\documentclass{article}

\usepackage{etoolbox}

\makeatletter
\patchcmd{\@makecaption}
  {\parbox}
  {\advance\@tempdima-\fontdimen2} 
  {}{}
\makeatother  

\usepackage[caption = false]{subfig}

\usepackage{anyfontsize}
\usepackage[a4paper]{geometry}
\usepackage{graphicx}
\usepackage[textwidth=8em,textsize=small]{todonotes}
    \usepackage{amssymb,euscript,amsfonts,mathrsfs,amscd}
\usepackage{amsmath}
\usepackage{natbib}
\RequirePackage{hyperref}

\def \bv {\textbf{v}}
\def \bZ {\textbf{Z}}
\def \bp {\textbf{p}}
\def \bx {\textbf{x}}
\def \bJ {\textbf{J}}

\title{An information-theoretic Phase I/II design for molecularly targeted agents that does not require an assumption of monotonicity~\footnote{This is an accepted version of the paper published in Journal of Royal Statistical Society: Series C (Applied Statistics), 62 (2), 2019, pp.1-21 \url{http://dx.doi.org/10.1111/rssc.12293}}}
\author{Pavel Mozgunov and Thomas Jaki \\ Medical and Pharmaceutical Statistics Research Unit, \\ Department of Mathematics and Statistics, \\ Lancaster University,
Lancaster,
UK.}
\date{ }

\begin{document}

\maketitle

\begin{abstract}
For many years Phase I and Phase II clinical trials were conducted separately, but there was a recent shift to combine these Phases. While a variety of Phase~I/II model-based designs for cytotoxic agents were proposed in the literature, methods for molecularly targeted agents (TA) are just starting to develop. The main challenge of the TA setting is the unknown dose-efficacy relation that can have either an increasing, plateau or umbrella shape. To capture these, approaches with more parameters are needed to model the dose-efficacy relationship or, alternatively, more orderings of the dose-efficacy relationship are required to account for the uncertainty in the curve shape. As a result, designs for more complex clinical trials, for example, trials looking at schedules of a combination treatment involving TA, have not been extensively studied yet. We propose a novel regimen-finding design which is based on a derived efficacy-toxicity trade-off function. Due to its special properties, an accurate regimen selection can be achieved without any parametric or monotonicity assumptions. We illustrate how this design can be applied in the context of a complex combination-schedule clinical trial. We discuss practical and ethical issues such as coherence, delayed and missing efficacy responses, safety and futility constraints. 
\end{abstract}

\section{Introduction} 
The primary objective of a Phase I trial is to identify the maximum tolerated regimen (dose, combination, schedule, etc.), that is the regimen corresponding to the highest acceptable toxicity probability $\phi$. Subsequently, a Phase II clinical trial with the objective to find a safe regimen corresponding to a pre-specified efficacy probability is planned. For many years these phases were conducted separately, but there was a recent shift to integrate both Phase I and Phase II clinical trials in a single study.  The integration of Phases allows to accelerate the development and reduces costs \citep{yin2012,wages} while more observations for both toxicity and efficacy endpoints become available. The goal of a Phase I/II clinical trial is to find the so-called optimal biologic regimen (OBR). There are different definitions of the OBR which depend on the context of the trial. The OBR can be defined as the safest regimen with a toxicity probability below the upper toxicity bound $\phi$ and a maximum efficacy above the lowest efficacy bound $\psi$ \cite[see e.g.][]{riviere2016}. It can, however, be challenging to find a single regimen having the safest toxicity and maximum efficacy simultaneously. In this case, the OBR can be alternatively defined as the regimen with the highest efficacy rate while still safeguarding patients (see e.g. \cite{wages2015}). In this work we use both definitions and call the regimen satisfying the first definition the \textit{optimal} regimen and the regimen satisfying the second definition the \textit{correct} regimen.

Historically, \cite{gooley} was one of the first to consider two dose-outcome models. Similar to later works on Phase I/II clinical trials \citep[e.g.][among others]{thall98,oquigley2001,braun2002,thall2004,yin2006}, this design is based on the paradigm `the more the better` - i.e. an agent has a greater activity but also a greater toxicity as the dose increases. This holds for cytotoxic agents, but can be violated for molecularly targeted agents (TA) which include hormone therapies, signal transduction inhibitors, gene expression modulators, apoptosis inducers, angiogenesis inhibitors, immunotherapies, and toxin delivery molecules, among others. For a TA either efficacy or toxicity curves can have a plateau \citep{morgan2003,postel2010,robert2014,xavier2014} or exhibit an umbrella shape \citep{conolly2004,lagarde2015}.

Several designs for either single agent or combination therapy trials that relax the assumption of monotonicity for the dose-efficacy relationship have been proposed in the literature: see e.g. \cite{polley2008,hoering2013,yin2013,zang2014, cai2014,wages2015,riviere2016} for a binary efficacy endpoint and \cite{hirakawa2012,winnie2015,yeung2017} for a continuous efficacy endpoint. The majority of current proposed designs are model-based and the selection of the OBR is governed either by (i) a trade-off function \citep[e.g.][]{thall2004,winnie2015,yeung2017} or (ii) a two-stage procedure in which the safe subset is firstly defined and the most efficacious dose is then estimated \citep[e.g.][]{thall98,yin2013,wages2015}. Regardless of the approach used, the number of parameters to be included in the underlying model increases considerably if the monotonicity assumption is relaxed \citep{cai2014,riviere2016}. The design proposed by \cite{wages2015} employs a one parameter model, but uses the idea of different orderings \citep{wages2011} to overcome the uncertainty about monotonicity. This idea  can be extended to a range of problems \citep[see e.g.][for cytotoxic drugs combination trials]{wages}, although the extension to more complex settings, for instance, combination-schedule trials, can also be challenging due a large number of possible orderings. Consequently, designs for such more complex trials have not been extensively studied yet. In contrast, an approach that does not use any parametric or monotonicity assumptions gives more flexibility and is a good candidate for both simple and complex Phase I/II clinical trials.

This work is motivated by an ongoing Phase I clinical trial at the Hospital Gustave Roussy  for patients with high-risk neuroblastoma, for which the authors contributed as statistical collaborators.  Neuroblastoma is the most frequent individual type of solid tumour in children \citep{steliarova2017international}. Although different chemotherapy regimens of increasing intensities have been evaluated, the 5-years overall survival remains around 50\% for the high risk group \citep{kreissman2013}. A recent pre-clinical study has suggested that the use of a particular immunotherapy targeting the disialoganglioside GD2 which is expressed in all neuroblastoma cells, in combination with conventional chemotherapy (etoposide and cisplatin) can improve the induction treatment.  The first part of the high-risk neuroblastoma treatment aims to reduce the tumour burden to facilitate surgery and subsequent treatments. Combinations of  the newly developed immunotherapy and chemotherapy are given under different schedules: 
\begin{itemize}
\item Immunotherapy  for 2 days after the chemotherapy ($S_1$)
\item Immunotherapy  for 3 days after the chemotherapy ($S_2$)
\item Immunotherapy  for 4 days together with the chemotherapy. Overlap $1$ day ($S_3$) 
\item Immunotherapy  for 4 days together with the chemotherapy. Overlap $2$ days ($S_4$)
\end{itemize} 
The combination of treatments is given for two cycles (three weeks). In each cycle the combination is given under one of schedules $S_1,\ldots,S_4$. Six different regimens are considered in the study and are given in Table \ref{tab:motivation}.
\begin{table}
\caption{\label{tab:motivation}The range of considered regimens in the motivating trial.}
\centering
\fbox{%
\begin{tabular}{c|cccccc}
Regimen & T$_1$ & T$_2$ & T$_3$ & T$_4$ & T$_5$ & T$_6$ \\
\hline
Cycle 1 & & $S_1$ & $S_2$ & $S_3$ & $S_3$ & $S_4$ \\
Cycle 2 & $S_1$ & $S_2$ & $S_2$ & $S_3$& $S_4$ & $S_4$ \\
\end{tabular}}
\end{table}
The toxicity outcome is evaluated by the end of the second cycle. The efficacy data is available after two more cycles (after six weeks from the start of the treatment) only. If a patient has experienced toxicity he will be treated off the protocol and the efficacy outcome cannot be observed. A maximum of $40$ patients will be recruited to the study and it is anticipated that two patients can be recruited each month. Consequently, it is expected that the next two patients are assigned a regimen before efficacy outcomes for the previous two patients are observed.

Clinician are certain that toxicity probabilities of $T_3,T_4$ and $T_5$ are greater than of $T_1,T_2$ and smaller than of $T_6$. However, toxicity probabilities of $T_3,T_4$ and $T_5$ cannot be put in the order of the increasing toxicity before the trial. Therefore, there are $6$ possibilities how regimens $T_1,\ldots,T_6$ can be ordered with respect to the toxicity probabilities
\noindent
 \begin{align}
  \begin{aligned}
1) \ T_1, T_2, T_3,T_4,T_5,T_6 \\
2) \ T_1, T_2, T_3, T_5, T_4,T_6 \\
3) \ T_1, T_2, T_4, T_3, T_5, T_6 \\
  \end{aligned}
  &&
  \begin{aligned}
4) \ T_1, T_2, T_4, T_5, T_3, T_6 \\  
5) \ T_1, T_2, T_5, T_3, T_4, T_6 \\
6) \  T_1, T_2, T_5,T_4, T_3, T_6 \\
  \end{aligned}
  \label{orderings}
 \end{align}
 
\noindent Additionally,  a plateau or umbrella shape for the efficacy is plausible due to the novel agent mechanism. This results in $48$ efficacy orderings to be considered so that traditional designs cannot be applied directly.

The aim of this paper is to propose a practical Phase I/II design that can be used for regimens with both monotonic and non-monotonic regimen-toxicity and regimen-efficacy relationships. We derive a novel trade-off criterion that governs the regimen selection during the study. The form of the proposed trade-off function is motivated by the recent developments in the theory of the weighted information measures \citep{entropy} and in the estimation on restricted parameter spaces \citep{mauro}. While model-based estimates can be used together with the proposed trade-off function, we show that a highly accurate optimal and correct regimens recommendations can be achieved without parametric assumptions due to the trade-off function's special properties. It is demonstrated how practical issues such as safety, futility or information about any pair of the regimens can be incorporated in the proposed design.
 
The rest of the paper is organised as follows. The information-theoretic approach, the trade-off function and its estimation are given in Section 2. The novel dose-finding design and its illustration are presented in Section 3. Safety and futility constraints are introduced in Section 4. Section 5 presents the application of the proposed design to the motivating trial. Further discussion is given in Section 6. 

\section{Methods}

\subsection{Trade-off function}

Following \cite{entropy}, let us consider a random variable $Y$ that takes one of three values corresponding to (i) `efficacy and no toxicity`, (ii) `no efficacy and no toxicity` and (iii) `toxicity`. Outcomes `toxicity and no efficacy` and `toxicity and efficacy` are combined as efficacy can only be observed when no toxicity occurs. Let $\bZ = \left[Z^{(1)},Z^{(2)},Z^{(3)}\right] \in \mathbb{S}^2$ be a random probability vector defined on the triangle
$$\mathbb{S}^2= \{\bZ: Z^{(1)}>0, \ Z^{(2)}>0, Z^{(3)}>0; \sum_{i=1}^3 Z^{(i)}=1 \}.$$
Random variables $Z^{(1)},Z^{(2)},Z^{(3)}$ correspond to probabilities of each of the three outcomes. As $\sum_{i=1}^3 Z^{(i)}=1$, the problem is, indeed, two-dimensional. A common question in a Phase I/II trial is `what is the probability vector $\bZ$'. We use a Bayesian framework to answer this question.

Assume that $\bZ$ has a prior Dirichlet distribution ${\rm Dir} (\bv + \bJ)$, $\bv= \left[v_1,v_2,v_3\right]^{\rm T} \in \mathbb{R}_{+}^{3}$ where $v_i>0$, $i=1,2,3$ and $\bJ$ is a $3$-dimensional unit vector. After $n$ realisations in which $x_i$ outcomes of $i$, $i=1,2,3$ are observed, the posterior probability density function $f_{n}$ of a random vector $\bZ^{(n)}$ takes the Dirichlet form
\begin{equation}
f_n(\bp|\bx) = \frac{1}{B(\bx+\bv + \bJ)}\prod_{i=1}^{3}p_i^{x_i+\nu_i}
\label{pdf}
\end{equation}
where $\bp = \left[p_1,p_2,p_3\right]^{\rm T}$, $\bx = \left[x_1,x_2,x_3 \right]$, $\sum_{i=1}^{3}x_i=n$, $0<p_i<1$, $\sum_{i=1}^{3}p_i=1$ and
$$B(\bx+\bv + \bJ) = \frac{\prod_{i=1}^3 \Gamma (x_i+v_i+1)}{\Gamma \left(\sum_{i=1}^{3}(x_i+v_i+1) \right) } $$
is the Beta-function and $\Gamma(x)$ is the Gamma-function.
The quantitative amount of information required to answer the estimation question can be measured by the Shannon differential entropy of $f_n$ \citep{cover2006} 
\begin{equation}
h(f_n) = - \int_{\mathbb{S}^2}  f_n(\bp|\bx) {\rm log} f_n(\bp|\bx) {\rm d}\bp
\label{entropy}
\end{equation}
with convention $0 \log 0 =0$. However, an investigator has a particular interest in regimens which are associated with desirable efficacy and acceptable toxicity probabilities and the measure (\ref{entropy}) does take this into account \citep{suhov2016}.

Assume that the OBR is defined as the regimen with probabilities of the three outcomes equal $\boldsymbol{\gamma}=\left[\gamma_1,\gamma_2,\gamma_3 \right] \in \mathbb{S}^2$ where $\gamma_1, \gamma_2, \gamma_3$ are the target probabilities of `no toxicity and efficacy`, `no toxicity and no efficacy` and `toxicity`, respectively. These target characteristics are to be defined by clinicians. Then, one would like to find the regimen which characteristics (probability of these events) as close as possible to the targets. Due to the specific interest in the regimens with characteristics close to $\boldsymbol{\gamma}$, let us consider a two-fold experiment in which, when the sample size is large, an investigator seeks to answers (i) what the probability vector is; while for the small sample size he is interested in (ii) whether the probability vector lies in the neighbourhood of $\boldsymbol{\gamma}$ so that subsequent patients can be allocated to regimens with the characteristics close to the OBR's ones.
The amount of the information required for this two-fold experiment can be measured by the \textit{weighted} Shannon differential entropy \citep{belis1968,clim2008,kelbert2015,suhov2016} of $f_n$ with a positive weight function $\phi_n(\bp)$
\begin{equation}
h^{\phi_n}(f_n) = - \int_{\mathbb{S}^2}  \phi_n (\bp) f_n(\bp|\bx) {\rm log} f_n(\bp|\bx) {\rm d}\bp.
\label{weighted_entropy}
\end{equation}
The weight function $\phi_n(\bp)$ emphasises an interest in the neighbourhood of $\boldsymbol{\gamma}$ rather than on the whole space. We will use a weight function of the Dirichlet form,

\begin{equation}
\phi_n(\bp) = C(\bx,\boldsymbol{\gamma},n) \prod_{i=1}^{3}p_i^{\gamma_i \sqrt{n}}
\label{weight}
\end{equation}
where the rate of $\sqrt{n}$ emphasises an interest in question (ii) for the small sample size only. Here $C(\bx,\boldsymbol{\gamma},n)$ is a constant that is chosen to satisfy the normalization condition $
 \int_{\mathbb{S}^2} \phi_n (\bp) f_n(\bp|\bx){\rm d}\bp =1.$

Due to ethical constraints and a limited sample size, an investigator is interested in the accurate estimation not for all regimens, but only for those with characteristics close to the target. This goal corresponds to questions (ii) alone. The quantitative measure of the information required to answer (ii) only  equals the difference of weighted $h^{\phi_n}(f_n)$ and standard $h(f_n)$ differential entropies \citep{entropy}. Denote the vector in the neighbourhood of which the probability density function $f_n$ concentrates as $n \to \infty$ by $\boldsymbol{\theta} = [\theta_1,\theta_2]^{\rm T}$, where $\theta_1$ is the probability of `efficacy and no toxicity`, $\theta_2$ is the probability of `no efficacy and no toxicity`, $1-\theta_1-\theta_2$ is the probability of toxicity with corresponding targets $\gamma_1$, $\gamma_2$, $1-\gamma_1-\gamma_2$. It was shown by \cite{entropy} that the difference of the entropies tends to

\begin{equation}
 \delta \left( \boldsymbol{\theta},\boldsymbol{\gamma}\right) := \frac{\gamma_1^2}{\theta_1} + \frac{\gamma_2^2}{\theta_2} + \frac{(1-\gamma_1-\gamma_2)^2}{1-\theta_1-\theta_2}-1, \ {\rm as} \ n \to \infty,
\label{result}
\end{equation}
the sum of three contributions corresponding to each of three events considered.

Although, the result (\ref{result}) is limiting, we propose to use this quantity as a measure of distance between $\boldsymbol{\theta}$ and $\boldsymbol{\gamma}$ and, consequently, as a trade-off function for selecting a regimen.  Clearly, $\delta(\cdot) \geq 0$, $\delta(\cdot)=0$ iff $\boldsymbol{\theta}=\boldsymbol{\gamma}$ and boundary values $\theta_i=0$, $i=1,2$ or $\theta_1+\theta_2=1$ correspond to infinite values $\delta(\cdot)$. The former property is argued to be  fundamental for restricted parameter spaces \citep{ait,logisticnorm2,mauro} and enables to avoid parametric assumptions. Terms in (\ref{result}) have the following interpretations:

\begin{itemize}
\item  When $\theta_1$ tends to $0$ the regimen is either inefficacious or (and) highly toxic. Then, the value of the trade-off function tends to infinity meaning that the treatment should be avoided.

\item When $\theta_2$ tends $0$ the regimen is either highly efficacious or (and) highly toxic. Then, this term penalises the high toxicity regardless of high efficacy as the trade-off function tends to infinity. This term prevents too quick escalation to highly toxic regimens.

\item When $1-\theta_1-\theta_2$ tends $0$, the regimen is associated with nearly no toxicity. However, the optimal regimen is expected to be associated with non-zero toxicity. Then, this terms drives dose allocation away from the underdosing regimens.
\end{itemize}
Note that all terms are dependent - an increase in one leads to decreases in others. The optimal value is attained at the point of target characteristics only. The denominators drive away the selection from inefficacious and highly toxic regimens and concentrate the allocation in the neighbourhood of the OBR.

The trade-off function (\ref{result}) depends on probabilities $\theta_1$ and $\theta_2$ while the goal of a Phase~I/II clinical trial is conventionally formulated in terms of  toxicity ($\alpha_t$) and efficacy ($\alpha_e$) probabilities and corresponding targets $\gamma_t$ and $\gamma_e$. Thus, we re-parametrise (\ref{result}) using $\theta_1= (1-\alpha_t) \alpha_e$, $\theta_2=(1-\alpha_t) (1-\alpha_e)$, $\gamma_1= (1-\gamma_t) \gamma_e$ and $ \gamma_2=(1-\gamma_t)(1-\gamma_e)$ and denote the trade-off function by $\delta \left( \alpha_t, \alpha_e, \gamma_t, \gamma_e \right)$. This measure can be computed for each of the $M$ regimens under study with  parameters
 $\alpha_{t,1},\ldots,\alpha_{t,M}$ and $\alpha_{e,1},\ldots,\alpha_{e,M}$, respectively.  Given  $\delta (\alpha_{t,i},\alpha_{e,i},\gamma_t,\gamma_e)$, $i=1,\ldots,M$, the target regimen $j$ satisfies \begin{equation}
\delta (\alpha_{t,j},\alpha_{e,j},\gamma_t,\gamma_e) = \min_{i=1,\ldots,M} \delta (\alpha_{t,i},\alpha_{e,i},\gamma_t,\gamma_e).
\end{equation}

\begin{figure}[!ht] 
\centering
\includegraphics[width=0.6\textwidth]{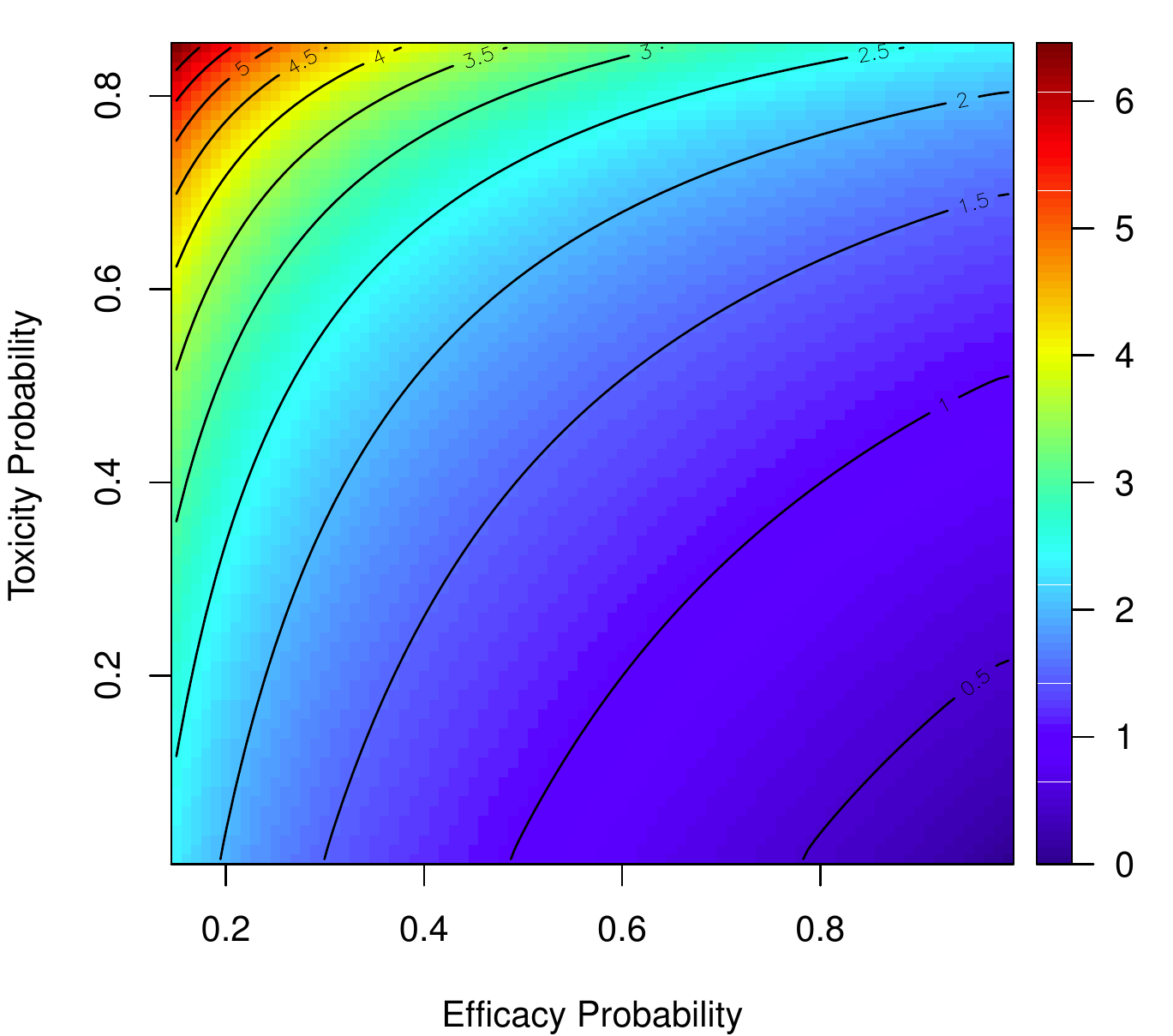}
\caption{\label{fig:loss_demo} Contours of the efficacy-toxicity trade-off function $\delta(\alpha_t,\alpha_e,\gamma_t,\gamma_e)$ for $\alpha_t \in (0,0.85)$ and $\alpha_e \in (0.15,1)$, $\gamma_t=0.01$, $\gamma_e=0.99$.}
\end{figure}

Efficacy-toxicity trade-off contours for different combinations of toxicity and efficacy probabilities and $\gamma_t=0.01$, $\gamma_e=0.99$ are given in Figure \ref{fig:loss_demo}.
 The contours of the trade-off function are concave. The most desirable point $\alpha_t=0.01, \ \alpha_e=0.99$ (right bottom corner) corresponds to the minimum of the trade-off function. As probabilities move away from this point the trade-off function grows with an increasing rate that can be seen by contours located closer to each other.

\subsection{Estimation}

The proposed trade-off function for regimen $i$ depends on unknown parameters $\alpha_{t,i}$ and $\alpha_{e,i}$. While these could be estimated using model-based approaches such as the continual reassessment method \citep{QPF90}, we can also consider each toxicity and efficacy probability independently as a Beta random variable estimated  directly from the observed number of toxicities/efficacies. This allows the design to avoid the monotonicity assumption which motivated this work initially.

Let us consider an estimator for regimen $i$. The toxicity probability has prior distribution $\mathcal{B}(\nu_{t,i}+1,\beta_{t,i}-\nu_{t,i}+1)$, $\nu_{t,i}, \beta_{t,i}>0$. After $n_i$ patients and $x_{t,i}$ toxicities, we obtain the Beta posterior $\mathcal{B}(x_{t,i}+\nu_{t,i}+1,n_i-x_{t,i}+\beta_{t,i}-\nu_{t,i}+1)$ which concentrates in the neighbourhood of $0<\alpha_{t,i}<1$. Similarly, given prior Beta parameters $\nu_{e,i}, \beta_{e,i}>0$ and $x_{e,i}$ efficacy responses the Beta posterior, $\mathcal{B}(x_{e,i}+\nu_{e,i}+1,n_i-x_{e,i}+\beta_{e,i}-\nu_{e,i}+1)$, for the efficacy probability can be found. The posterior modes of these Beta distributions are
\begin{equation}
\hat{p}_{t,i}^{(n_i)} = \frac{x_{t,i}+\nu_{t,i}}{n_i+\beta_{t,i}} \ \ \ \ \ \ {\rm and} \ \ \ \ \ \ \hat{p}_{e,i}^{(n_i)} = \frac{x_{e,i}+\nu_{e,i}.}{n_i+\beta_{e,i}},
\label{est1}
\end{equation}
respectively. We then use the `plug-in` estimator $\delta (\hat{p}_{t,i}^{(n_i)},\hat{p}_{e,i}^{(n_i)},\gamma_t,\gamma_e) = \hat{\delta}_i^{(n_i)}$ as the criterion that governs the selection of a subsequent regimen to be studied in a Phase~I/II regime-finding trial.

We would like to emphasize that the trade-off function (\ref{result}) is derived in the general form without assuming independence between toxicity and efficacy. While the estimators (\ref{est1}) does not account for the interaction between toxicity and efficacy, the derived trade-off function does. The independent estimates are used to avoid a monotonicity assumption, but the trade-off function in the derived form allows any  estimator to be applied.

\section{Regimen-finding algorithm}

\subsection{Non-randomised design}

We propose the following novel regimen-finding design that does not require the assumption of monotonicity for trials looking to identify a regimen with toxicity probability $\gamma_t$ and efficacy probability $\gamma_e$. Assume that $N$ patients and $M$ regimens are available in a trial. Patients are assigned sequentially cohort-by-cohort, where a cohort is a small group of typically $1$ to $4$ patients. Let $\hat{\delta}^{(n_i)}_i$ be the `plug-in` estimator for regimen $i$ after $n_i$ patients have been treated on regimen $i$. The procedure starts from $\hat{\delta}^{(0)}_i$ and the starting regimen depends on the prior only. After $n_i$ patients on regimen $i$ ($i=1,\ldots,M$) the next cohort of patients is allocated to a regimen $m$ such that
\begin{equation}
\hat{\delta}^{(n_m)}_m= \min_{i=1,\ldots,M} \hat{\delta}^{(n_i)}_i.
\label{criterion}
\end{equation}
The method proceeds until the maximum number of patients, $N$, have been treated. We adopt the regimen selected after $N$ patients as the final recommended regimen. 

Note that the estimators in (\ref{est1}) require a vector of prior probabilities for each regimen to be specified before the trial. This choice implies a specific ordering of the regimens before trial data are available and can be considered as an initial ordering. However, as the trial progresses this ordering can change. We refer to the proposed design as `WE` which stands for the weighted entropy which motivated the criterion.

\subsection{Randomized design}
For the WE design all previous outcomes are taken into account for the current selection decision. In this case designs can `lock-in` which means that one regimen would be tested regardless of further outcomes and the true optimal regimen can be never tested. It might happen due to certain sequences of observations in combination with the choice of the prior. Thus, one can benefit from the allocation rule based on a randomization \citep{thall2007,wages2015}. We propose randomization within a safety set with probabilities proportional to the inverse of the trade-off function. For ethical considerations we restrict randomization to the two best regimens only. Formally, assume that regimen $m$ is the estimated best regimen (has the minimum value of $\hat{\delta}^{(n_m)}_m$) and $j$ is the second best regimen (i.e. has the second smallest value of $\hat{\delta}^{(n_j)}_j$). Then, randomization probabilities  $\hat{w}_i^{(n_i)}$ for a regimen $i=1,\ldots,M$ are
\begin{equation}
 \hat{w}_i^{(n_i)} =
 \begin{cases} \frac{1/{\hat{\delta}^{(n_i)}_i}}{1/\hat{\delta}^{(n_m)}_m+1/\hat{\delta}^{(n_j)}_j}, \ {\rm if} \ i=m,j \ {\rm and} \ \hat{\delta}^{(n_m)}_m,\hat{\delta}^{(n_j)}_j \neq 0. \\
  1, \ {\rm if} \ \hat{\delta}^{(n_i)}_i = 0. \\
   0,  \ {\rm otherwise.} \\
 \end{cases}
\end{equation}
The method proceeds until the maximum number of patients, $N$, have been treated. The regimen $m$ satisfying (\ref{criterion}) is adopted as the final recommendation. We refer to this design as `WE(R)`. This randomization technique allows to get more spread allocation, while assigning only few patients to suboptimal regimens. We will focus on these two allocation rules only although other alternatives are possible. For instance, one might assign the first patients using randomization and the rest using allocation to the `best` as suggested by \cite{wages2015}.

\subsection{Delayed responses}
So far, it was implied that efficacy responses are available at the same time as the toxicity information is. However, in practice this is unlikely to be true. While toxicity is usually quickly ascertainable, the efficacy endpoint may take longer to be observed \citep{riviere2016} and waiting for both endpoints increases the length of a trial substantially. However, the proposed criterion still can be applied in the trial with delayed responses as $\hat{p}_{t,i}^{(n_i)}$ and $\hat{p}_{e,i}^{(n_i)}$ can be estimated based on a different number of observations. Consequently, the design can proceed before the full response is observed and would only use the information available at the time  the regimen for the next cohort is selected.

For instance, in the motivating trial it takes twice as long to observe the efficacy outcome than the toxicity outcome (six weeks versus three weeks). To conduct the trial in a timely manner, the next  cohort of patients is expected to be assigned based on the toxicity data only for the previous cohort and both toxicity and efficacy data for earlier patients. For instance, the recommendation for the third cohort is based on both toxicity and efficacy outcomes for cohort $1$ but only the toxicity outcomes for cohort $2$. The proposed design accommodates this by computing toxicity and efficacy estimates based on different numbers of observations, but on all the available information up to the time of the next patient allocation. Note that if an efficacy outcome is available earlier, it can (and should) be included in the design. Doing so can improve the performance of the design as illustrated in the supplementary materials. Note also that there may be situations in which auxiliary information about efficacy is available (e.g. through a short-term endpoint). Accounting for this information is beyond the scope of this manuscript.

\subsection{Coherence}
In practice, a clinician might be very cautious about further escalation if a toxicity was observed in the previous cohort. For this reason, we force the WE designs to satisfy principles of the coherent escalation/de-escalation \citep{coherent} with respect to known orderings. Assume that there are $S$ known monotonic partial orderings. Denote the position of a regimen for cohort $i$ in the partial ordering $s$ by $_{[s]}T^{(i)}$ and the sum of the corresponding toxicity outcomes by $Q^{(i)}$. Then the coherent escalation  means that,
\begin{equation}
\mathbb{P} \left(_{[s]}T^{(i)}-_{[s]}T^{(i-1)}>0 | Q^{(i-1)} \geq q \right) = 0, \ s \in S
\label{coh:escalation}
\end{equation}
where $q$ is a threshold number of toxicities after which the escalation should be prohibited. The coherent de-escalation means that
\begin{equation}
\mathbb{P} \left(_{[s]}T^{(i)}-_{[s]}T^{(i-1)}<0 | Q^{(i-1)} < q \right) = 0, \ s \in S.
\label{coh:deescalation}
\end{equation}
For instance, in the motivating example the following partial orderings are known
$$ \alpha_{t,1} \leq \alpha_{t,2} \leq \alpha_{t,3}\leq \alpha_{t,6} $$
\begin{equation}
\alpha_{t,1} \leq \alpha_{t,2} \leq \alpha_{t,4} \leq \alpha_{t,6}  
\label{orderings2}
\end{equation}
$$ \alpha_{t,1} \leq \alpha_{t,2} \leq \alpha_{t,5} \leq \alpha_{t,6}.  $$
It follows that if more than $q$ toxicity outcomes are observed for $T_3$ the next cohort can be still allocated to $T_4$. Moreover, this coherence principles are used to incorporate the information about pairs of regimens that can be ordered (for example $T_1$ and $T_2$). While it is not taken into account in the estimation step of the original design, it is reflected in allocation restrictions.

\subsection{Illustration \label{sec:illustration}}
Below, we demonstrate the performance of the non-randomized WE design in the context of the motivating trial. The major challenge of the motivating trial is the uncertainty in the regimen-toxicity relation for $T_3,T_4,T_5$ which results in six possible toxicity orderings (\ref{orderings}). Furthermore, for each of these toxicity orderings either a monotonic, plateau or umbrella regimen-efficacy relationship can be expected. Despite this complex setting, the WE design can be applied as both toxicity and efficacy endpoints are binary.

We consider the regimen-finding clinical trial with $M=6$ regimens, $T_1,\ldots,T_6$, $N=36$ patients and cohort size $c=2$. The regimens are ordered (on the basis of clincians believe) with increasing toxicity and efficacy. The trial is to be started at regimen $T_1$ and no regimen-skipping is allowed.  The coherence parameter is fixed to be $q=1$ and the escalation/de-escalation is required to be coherent with respect to known partial orderings (\ref{orderings2}). Following the motivating trial, toxicity is evaluated after two cycles of treatment (three weeks) and efficacy data is available after four cycles (six weeks) only. Since it is expected to recruit one patient per month,  we assume that the next patient is assigned after the toxicity outcome is available for the previous patient. Moreover, efficacy is only observed for patients without toxicity.

True probabilities of toxicity and efficacy are
$\alpha_t = [0.05,0.10,0.45,0.15,0.30,0.55]^{\rm T} $
and
$\alpha_e = [0.10,0.40,0.70,0.70,0.70,0.70]^{\rm T}.$
This scenario corresponds to a plateau in the regimen-efficacy relationship starting at $T_3$ and to the misspecified ordering of toxicities $-$ regimen $T_3$ is more toxic than regimens $T_4$ and $T_5$. We study the ability of the WE design to recommend the \textit{optimal} and \textit{correct} regimens. The safest regimen with a toxicity probability below the upper toxicity bound $\phi$ and a maximum efficacy above the lowest efficacy bound $\psi$ is called \textit{optimal}  and the regimen with the highest efficacy rate while still safeguarding  patients (irrespective of it also having lowest toxicity) is called \textit{correct}. Then, for a maximum toxicity probability bound ${\phi}=0.35$ and a minimum efficacy probability bound $\psi=0.20$, regimen $T_4$ is the optimal one and regimens $T_4$ and $T_5$ are correct regimens. To ensure that a regimen with the same efficacy but lower toxicity is preferred over one with higher toxicity we set $\gamma_t=0.01$ and, similarly, we set $\gamma_e=0.99$ to prefer a regimen with a higher efficacy if the toxicity is the same. To see that $T_4$ is indeed the OBR using these targets, one can compute the true values of the trade-off function [9.48,1.77,1.59,\textbf{0.67},1.03,2.16]$^{\rm T}$. The minimum among all regimes corresponds to $T_4$ as desired and the second smallest value corresponds to the correct regimen $T_5$.

\begin{figure}[!ht] 
\centering
 \includegraphics[width=0.85\textwidth]{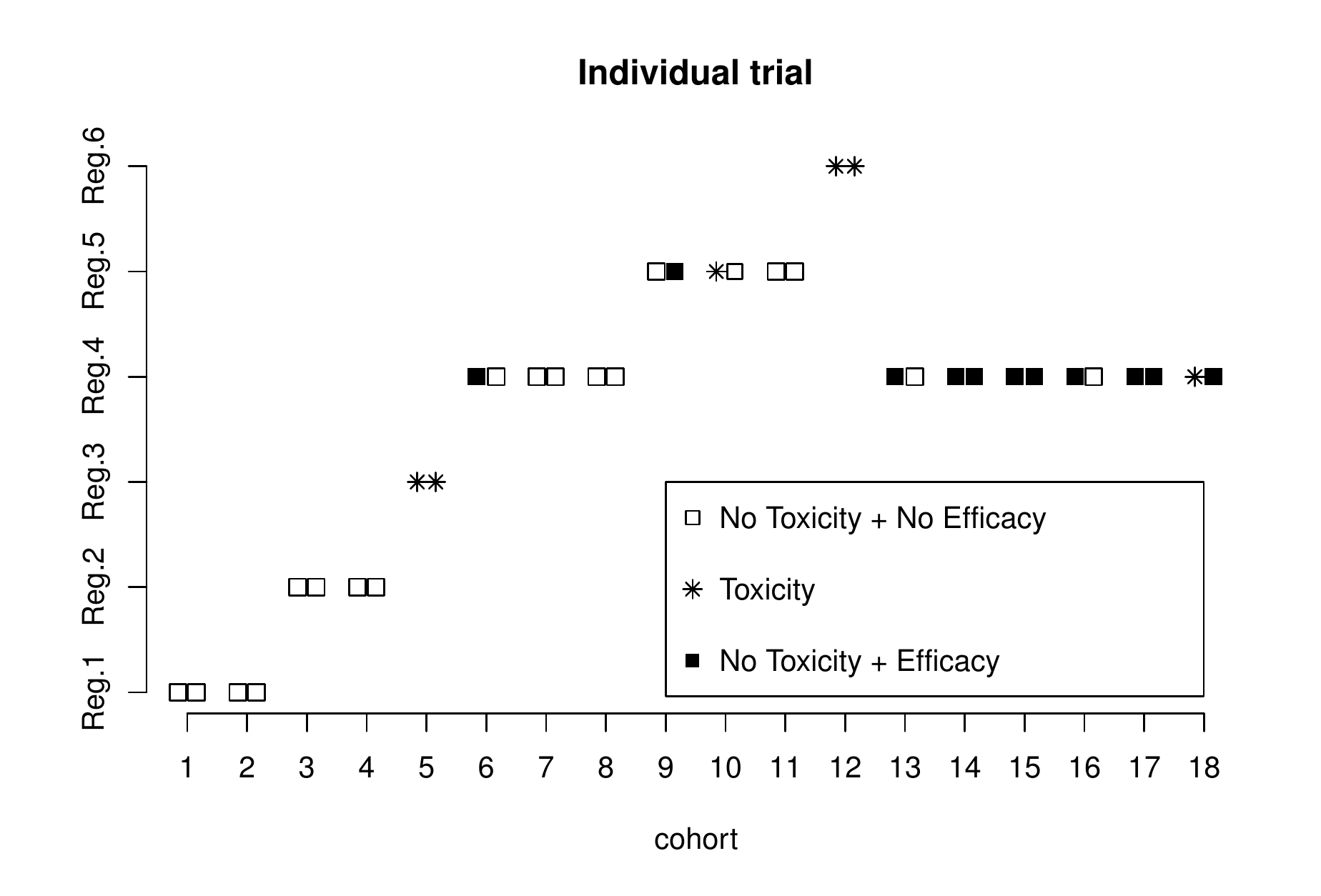}
 \caption{Allocation of $18$ cohorts in the invididual trial. $\square$ corresponds to `no toxicity and no efficacy`, $\star$ to `toxicity` and  $\blacksquare$ to `no toxicity and efficacy` responses. \label{fig:individ}}
\end{figure}

Prior parameters of toxicity $$\hat{p}^{(0)}_t= [0.10,0.175,0.25,0.325,0.40,0.475]^{\rm T},$$ efficacy $$\hat{p}^{(0)}_e= [0.60,0.65,0.70,0.75,0.80,0.85]^{\rm T}$$ and  $\beta_{t,i}=\beta_{e,i}=1$, $i=1,\ldots,6$  are chosen. Note that the purpose of this priors is to specify in which order the regimens are trialled. The allocation of $18$ cohorts in a single simulated trial is given in Figure \ref{fig:individ}. 
Note that `no toxicity and efficacy` and `no toxicity and no efficacy` outcomes in Figure \ref{fig:individ} can be observed after four cycles only. Until an efficacy outcome is available the design uses information about `no toxicity` only.

The allocation starts at $T_1$ with no toxicities. As no efficacy has been observed yet and $T_1$ has `promising` prior efficacy probability, it is selected again. Following no efficacies for cohort 1, cohort $3$ is assigned to $T_2$ at which no patients have toxicities. This leads to selecting $T_2$ again until the efficacy data are available. After no toxicity (cohort 4) and no efficacy (cohort $3$), cohort 5 is assigned to $T_3$ for which both patients experience toxicities. As there is uncertainty whether $T_3$ is more toxic than $T_4$ and (or) $T_5$, cohort 6 is assigned to $T_4$. After no toxicity outcomes are observed, cohort 7 is allocated to regimen $T_4$ again. Due to no toxicity (cohort 7) and one efficacy (cohort 6), regimen $T_4$ is chosen for cohort 8 as well. However, after no efficacy for cohort 7, the design escalates to regimen $T_5$. Again, as no toxicity outcomes are observed for cohort 9, cohort 10 is assigned to $T_5$ too at which one patient experiences a toxicity. However, by the time cohort 11 is allocated, an efficacy outcomes for cohort 9 becomes available and the allocation remains at regimen $T_5$. As no further efficacy has been observed for regimen $T_5$, design escalates further to regimen $T_6$ for which two toxicities are observed. Then, the design de-escalates to the optimal regimen $T_4$ for which one efficacy and no toxicity has been previously observed (against 1 efficacy and 1 toxicity for regimen $T_5$). All the consequent patients (up to cohort 18) are assigned to the optimal regimen $T_4$ which is finally recommended in the trial. Clearly, a delayed efficacy response requires two cohort to be assigned to each dose conditionally on `no toxicity`. It leads to fewer patients at the OBR, but also to more reliable recommendation due to better exploration of the available regimes. 

\begin{figure}[!ht] 
\centering
 \includegraphics[width=0.85\textwidth]{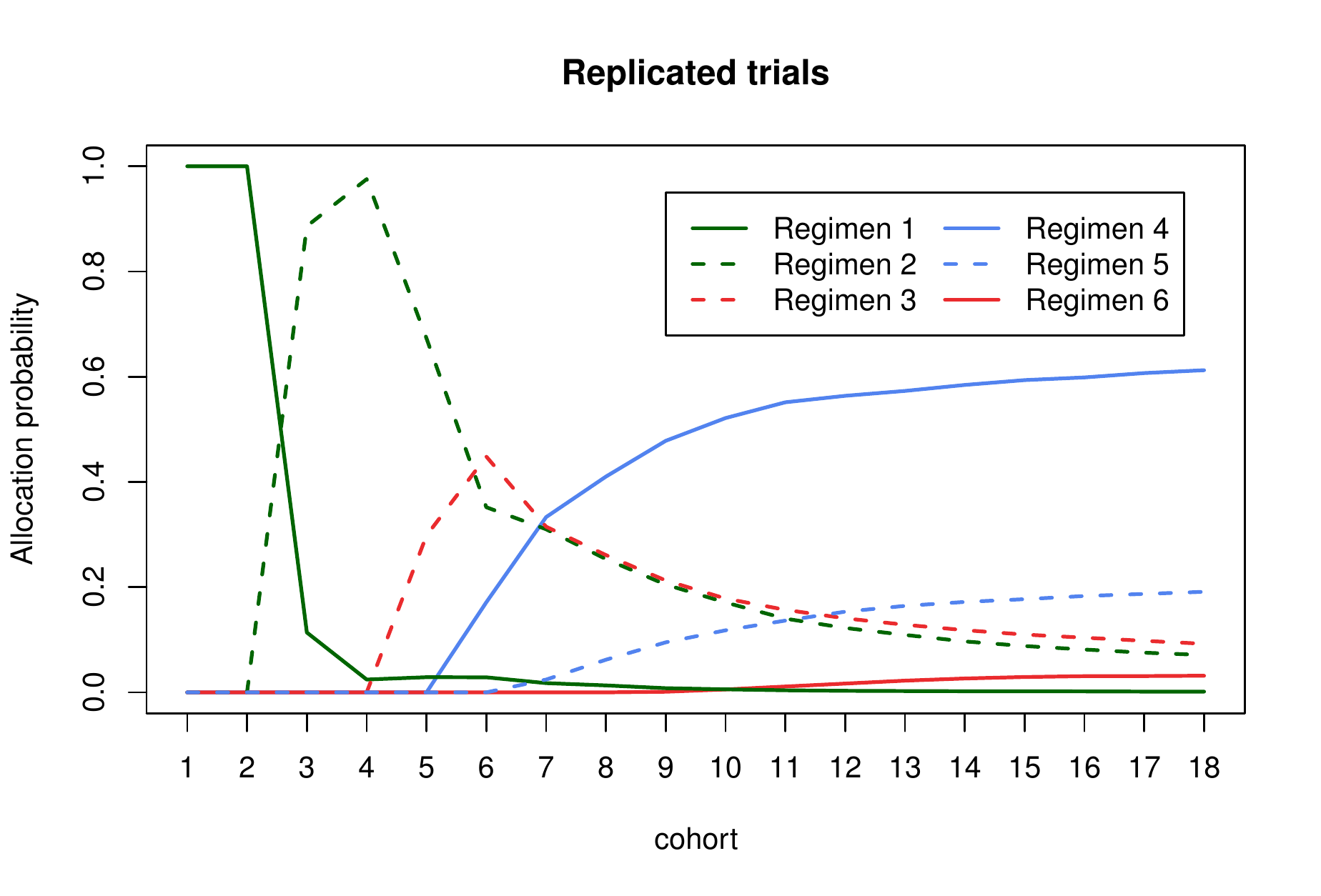}
 \caption{Probabilities to allocate each of $18$ cohorts to $T_1$ (solid green), $T_2$ (dashed green), $T_3$ (dashed red), $T_4$ (solid blue), $T_5$ (dashed blue) and $T_6$ (solid red) in the motivating trial setting. The optimal regimen is Regimen $T_4$ and the correct regimens are Regimens $T_4$ and $T_5$. Results are based on $10^6$ replications. \label{fig:probability}}
\end{figure}

While the regimen-finding algorithm in an individual trial is considered above, allocation probabilities for each cohort in $10^6$ replicated trials are given in Figure \ref{fig:probability}. 
Using the WE design, first and second cohorts are to be assigned to the first regimen with probability 1. As it illustrated above, the design stays at $T_1$ if no toxicity was observed (due to a `promising` efficacy) or if a toxicity is observed  (due to the coherence principle). The probability to allocate a cohort to the optimal regimen $T_4$ starts to increase after cohort $5$ and reaches nearly 60\%  for cohort 18. Considering the probability of regimen recommendations, we compare the performance to an equal allocation of $6$ patients per each regimen. For the equal allocation, the trade-off functions for each regimen are estimated at the end of the trial and the dose with a smallest value is recommended. The optimal regimen $T_4$ is recommended in 62.5\% of trials and the correct regimen $T_5$ in 18.6\% of trials  by the WE design against 31\% and 29\% by an equal allocation, respectively. It means that the proposed design recommends one of the correct regimens in more than 80\% of the trials and clearly favours the optimal one.

Overall, the proposed design appears to be able to recommend the optimal regimen with high probability and the escalation/de-escalation algorithm in the individual trial is intuitive. A comprehensive study comparing the proposed method to alternative approaches and across different scenarios is given in Section \ref{sec:numerical}. \\

\section{Ethical constraints}
Not borrowing information across regimens is a key feature of the proposed design. However, some regimens might have high toxicity and/or  low efficacy. Then, a design can result in a high (small) number of toxicity (efficacy) responses or in an unsafe/inefficacious recommendation. For ethical reasons it is required to control the number of patients exposed to such regimens and two time-varying constraints are introduced.

\subsection{Safety constraint}
Absence of the monotonicity assumption for toxicity makes the problem of the highly toxic regimen selection even more crucial. A conventional (constant) safety constraint \citep[e.g. as in][]{riviere2016} cannot be applied because no parametric model is used \citep{entropy}. On the one hand, a reliable safety constraint should give the hypothetical possibility to test all regimens if data suggests so. On the other hand, a recommendation should be made with a high confidence in safety. A time-varying safety constraint meets both of these requirements. A regimen $i$ is safe if after $n_i$ patients 
 \begin{equation}
       \int_{\phi^*}^{1}f_{t,i}^{(n_i)}(p) {\rm d}p \leq \zeta^{(n_i)}
       \label{safety}
       \end{equation}
where $f_{t,i}^{(n_i)}$ is the Beta posterior density function of the toxicity probability, $\phi^*$ is the toxicity threshold and $\zeta_{n_i}$ is the probability that controls overdosing. As information increases, we gain confidence about a regimen's safety and hence consider the constraint that becomes more strict as the trial progresses. We therefore use a  non-increasing function of $n_i$ for $\zeta^{(n_i)}$. We choose $\zeta^{(0)}=1$ initially to allow all regimens to be tested while the final recommendation is made with probability $\zeta_{N}$. Subsequently we use a linear decreasing function
$
 \zeta^{(n_i)}= max(1-r_tn_i,\zeta_{N})
$
where $r_t>0$. These safety constraint parameters can either be specified by experts or alternatively calibrated with respect to a trial's goals using simulations. 

\subsection{Futility constraint}
The same reasoning is applied to a time-varying futility constraint. Regimen $i$ is efficacious if after $n_i$ patients
 \begin{equation}
       \int_{\psi^*}^{1}f_{e,i}^{(n_i)}(p){\rm d}p \geq \xi^{(n_i)}
       \label{futility}
       \end{equation}
where $f_{e,i}^{(n_i)}$ is the Beta posterior density function of the efficacy probability, $\psi^*$ is the efficacy threshold and $\xi^{(n_i)}$ is the controlling probability. This probability is an increasing function of $n_i$ and the recommendation is made with probability $\xi_{N}$. We use a linear increasing function
$
 \xi^{(n_i)}= min(r_en_i,\xi_{N})
$
where $r_e>0$.

\section{Numerical results \label{sec:numerical}}
\subsection{Setting}
In this section we study the performance of the proposed design in the context of the motivating trial under various different scenarios. Following the motivating trial, we consider $M=6$ regimens and $N=36$ patients which are enrolled in cohorts of $c=2$. The setting as stated in Section \ref{sec:illustration} remains unchanged. The upper toxicity and the lowest efficacy bounds are $\phi=0.35$ and $\psi=0.20$, respectively. The goal is to study the ability of the WE design to identify the \textit{optimal} and \textit{correct} regimens as defined above.
 
\begin{figure}[!ht] 
\centering
 \includegraphics[width=1\textwidth]{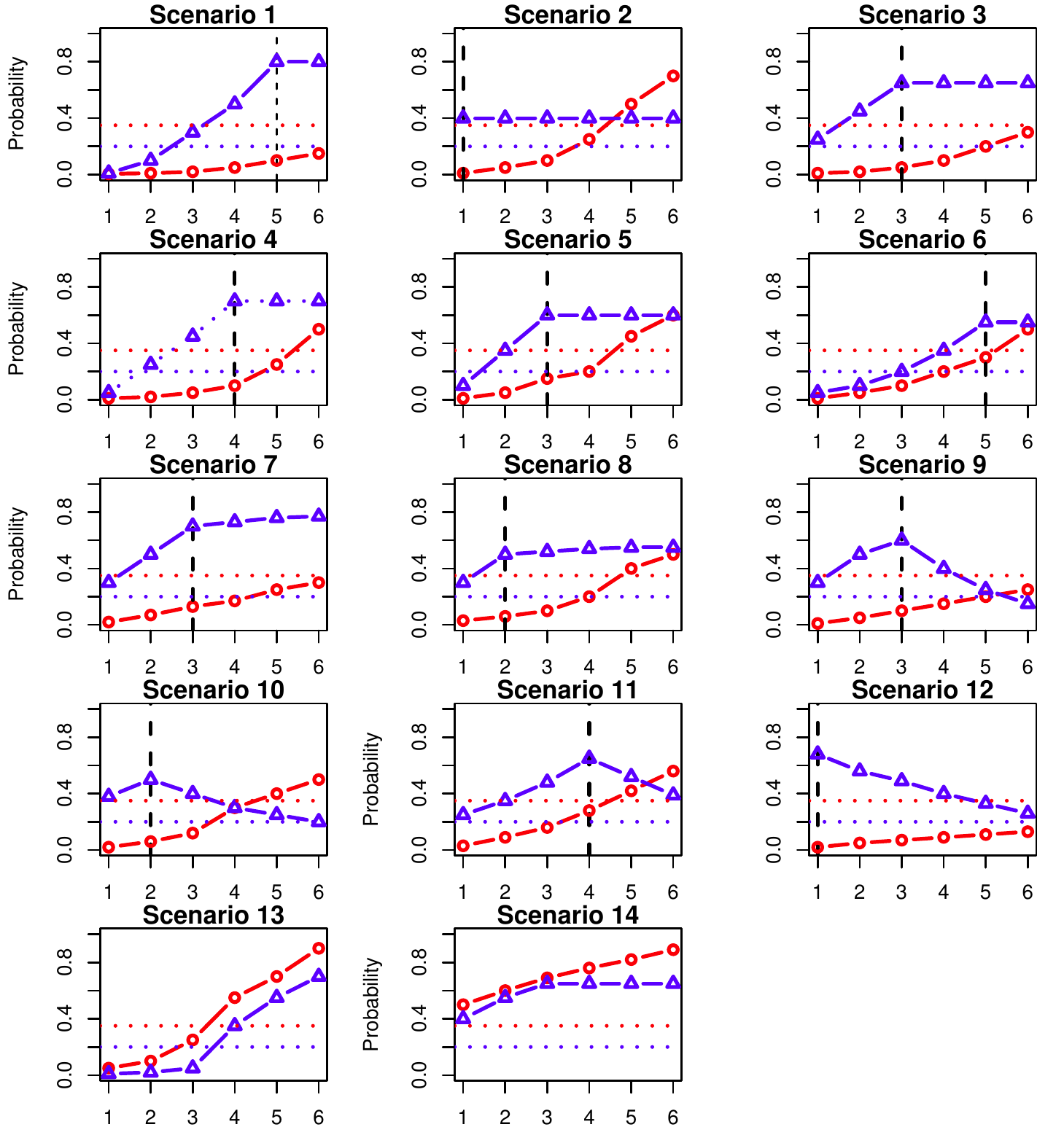}
 \caption{Eight plateau regimen-efficacy scenarios (1-8), four umbrella regimen-efficacy scenarios (9-12) and two scenarios with no correct regimens (13-14) in the trial with $M=6$ regimens. Toxicity  and efficacy probabilities are marked by red cicles and blue triangles, respectively. Red horizontal line corresponds to the upper toxicity bound $\phi=0.35$ and blue dashed horizontal line coresponds to the lowest efficacy bound $\psi=0.20$. Dashed black vertical line corresponds to the optimal regimen.\label{fig:scenarios1}}
\end{figure}

The major challenge of the motivating trial is the uncertainty in both regimen-toxicity and regimen-efficacy relations. This increases the number of possible scenarios to be investigated enormously. Therefore, we start by defining scenarios on which the  assessment will be based. Firstly, we specify 14 scenarios with increasing regimen-toxicity relations and different shapes of regimen-efficacy curves (Figure \ref{fig:scenarios1}): eight plateau regimen-efficacy scenarios (1-8) by \cite{riviere2016}, four umbrella shaped scenarios (9-12) by \cite{wages2015}, and two scenarios with no optimal and correct regimens (13-14, due to inefficacy and toxicity, respectively).
These scenarios were originally used to test the WE approach for Phase I/II dose finding designs with a single TA and compare its performance to the methods by \cite{wages2015} and \cite{riviere2016}. The results of this evaluation are given in the supplementary materials. Secondly, to allow for the uncertainty in the toxicity ordering, we consider six permutations of each scenario with respect to toxicity orderings (\ref{orderings}). For instance, six permutations of scenario 1 are given in Table \ref{tab:newscenarios}.
\begin{table}
   \caption{Six permutations of scenario 1. The optimal regimen is in \textbf{bold} and correct regimens are \underline{underlined}. \label{tab:newscenarios} }
  \fbox{%
\begin{tabular}{ccccccc}
   & $T_1$ &   $T_2$ &  $T_3$ &   $T_4$ &   $T_5$ & $T_6$ \\
   \hline
Scenario 1.1 & (.005;.01) & (.01;.10) & (.02;.30) & (.05;.50) & \underline{\textbf{(.10;.80)}} & \underline{{(.15;.80)}}\\
Scenario  1.2 & (.005;.01) & (.01;.10) & (.02;.30) & \underline{\textbf{(.10;.80)}} & (.05;.50) & \underline{{(.15;.80)}} \\
Scenario   1.3 & (.005;.01) & (.01;.10) & (.05;.50) & (.02;.30)  & \underline{\textbf{(.10;.80)}} & \underline{{(.15;.80)}}\\
Scenario 1.4 & (.005;.01) & (.01;.10) & \underline{\textbf{(.10;.80)}}  & (.02;.30)  & (.05;.50) & \underline{{(.15;.80)}}\\
Scenario  1.5 & (.005;.01) & (.01;.10) & (.05;.50) & \underline{\textbf{(.10;.80)}} &  (.02;.30) & \underline{{(.15;.80)}}\\
Scenario 1.6 & (.005;.01) & (.01;.10) & \underline{\textbf{(.10;.80)}}  & (.05;.50) & (.02;.30) & \underline{{(.15;.80)}}\\
    \hline
 \end{tabular}}
\end{table}
Overall, this results in $84$ scenarios that cover a large variety of possibilities and allows the proposed design to be assessed in the setting of the motivating trial adequately. In the analysis we focus on (i) the proportion of optimal/correct recommendations, (ii) the average number of toxic responses, (iii) the average number of efficacy responses. The study is performed using \texttt{R} \citep{Rcore} and 10,000 replications for each scenario. We compare the characteristics to an extended `WT` design by \cite{wages2015} whose specification is given below.

\subsection{Design specification}

As in the illustrative example above, we use a target toxicity of $\gamma_t=0.01$ and a target efficacy of $\gamma_e=0.99$. This implies that an investigator targets the most efficacious yet  the least toxic regimen.  Due to the partially unknown toxicity ordering, the design was restricted to satisfy the coherence conditions (\ref{coh:escalation})-(\ref{coh:deescalation}) with respect to partial orderings and $q=1$. The randomized design is presented here as it has been shown in the evaluations of single agent trials (see supplementary materials) to be have more potential benefits when there is more than one correct regimen expected in the trial.

Parameters $\beta_{t,i}=\beta_{e,i}=1$ of the prior Beta distributions in (\ref{est1}) are chosen for all regimens $i=1,\ldots,6$ to emphasize a limited amount of information available to clinicians. Parameters $\nu_{t,i}$ and $\nu_{e,i}$ (which coincide with prior toxicity efficacy probabilities for chosen values of $\beta_{t,i}, \beta_{e,i}$) are calibrated such that the optimal regimen can be found in various different scenarios with high probability. Note that the design is fully driven by the values of the trade-off function and, therefore, there are two requirements to the prior parameters. The prior should dictate to start the trial at the lowest regimen $T_1$ and the design should follow the escalation order of regimens specified by clinicians (with no regimen-skipping). To preserve a gradual escalation, the prior should assume that the higher regimens have greater efficacy, but also greater toxicity. To restrict the number of prior parameters to be calibrated, we would assume a linear increase in prior toxicity and efficacy probabilities. Through extensive calibration, prior vectors of toxicity probabilities $\hat{\bp}_t^{(0)} = \left[0.10, 0.14, 0.18, 0.22, 0.26, 0.30 \right]^{\rm T}$ and efficacy probabilities $\hat{\bp}_e^{(0)} = \left[0.60, 0.62, 0.64, 0.66, 0.68, 0.70 \right]^{\rm T}$ were chosen for subsequently analysis in all scenarios. Note that despite the increasing prior probabilities, the ordering of the regimes is not fixed and can change as the trial progresses.

Regarding the ethical constraint, parameters of the safety constraint $\zeta_{N}=0.30, r_t=0.02$, $\phi^*=0.4$  and of the futility constraint $\xi_{N}=0.50, r_e=0.05, \psi^* = 0.35$ were calibrated. Further guideline on the prior parameters choice together with the calibration of safety and futility constraints are given in supplementary materials.

We compare the performance of the novel approach to the extension of the WT design by \cite{wages2015}. We extend the original WT design to allow the randomization between toxicity orderings. Following \cite{wages2015}, a trial with a monotonic regimen-toxicity relation and $6$ regimens is associated with $11$ efficacy orderings: one strictly monotonic, five cases of a plateau location and 5 cases of a umbrella peak. Similarly, one can deduce all possible efficacy orderings associated with each toxicity ordering. For instance, for the toxicity ordering $T_1, T_2, T_3,T_5,T_4,T_6$ we specify
\begin{enumerate}
\item 0.10, 0.20, 0.30, 0.50, 0.40, 0.60 (monotonic with respect to regimen-toxicity) 
\item 0.20, 0.30, 0.40, 0.60, 0.50, 0.60 (plateau starting at $T_4$)
\item 0.30, 0.40, 0.50, 0.50, 0.60, 0.40 (peak at $T_5$)
\item 0.20, 0.30, 0.40, 0.60, 0.50, 0.50 (peak at $T_4$)
\item 0.40, 0.50, 0.60, 0.40, 0.50, 0.30 (peak at $T_3$)
\item 0.50, 0.60, 0.50, 0.30, 0.40, 0.20 (peak at $T_2$)
\item 0.60, 0.50, 0.40, 0.20, 0.30, 0.10 (peak at $T_1$)
\end{enumerate}  
Note that other orderings as a `plateau starting at $T_1$/$T_2$/$T_3$/$T_5$` are already included in the first regimen-toxicity case. Applying the same procedure to the rest of toxicity orderings in (\ref{orderings}) leads to $48$ efficacy orderings. The design proceeds as follows. Firstly, given the available data, one of $6$ toxicity orderings is selected as proposed by \cite{wages2011}. Secondly, one of $48$ efficacy orderings is chosen as in the original design. The parameters of the designs are chosen as in the original specification by \cite{wages2015_code} with an exception of using cohort size $c=2$  and $80\%$ confidence intervals for stopping rules. Note that the WT design waits for both toxicity and efficacy responses to allocated the next cohort and assumes that an efficacy outcome is observed regardless the toxicity outcome.

Different variation of the extended WT design were also explored.  A reduced number of efficacy orderings (plateau cases only) were investigated, but no significant difference was found,  and the specification with the full number of orderings was found to be more robust. Additionally, the `conditional` choice of the efficacy orderings was also studied. Using this approach, once the toxicity orderings is selected, that choice of the efficacy orderings is restricted to 11 orderings with respect to the toxicity profile. However, it was found to result in  less accurate optimal and correct regimen recommendations across all scenarios.

\subsection{Operating characteristics}
\begin{figure}[!ht] 
\centering
\includegraphics[width=1\textwidth]{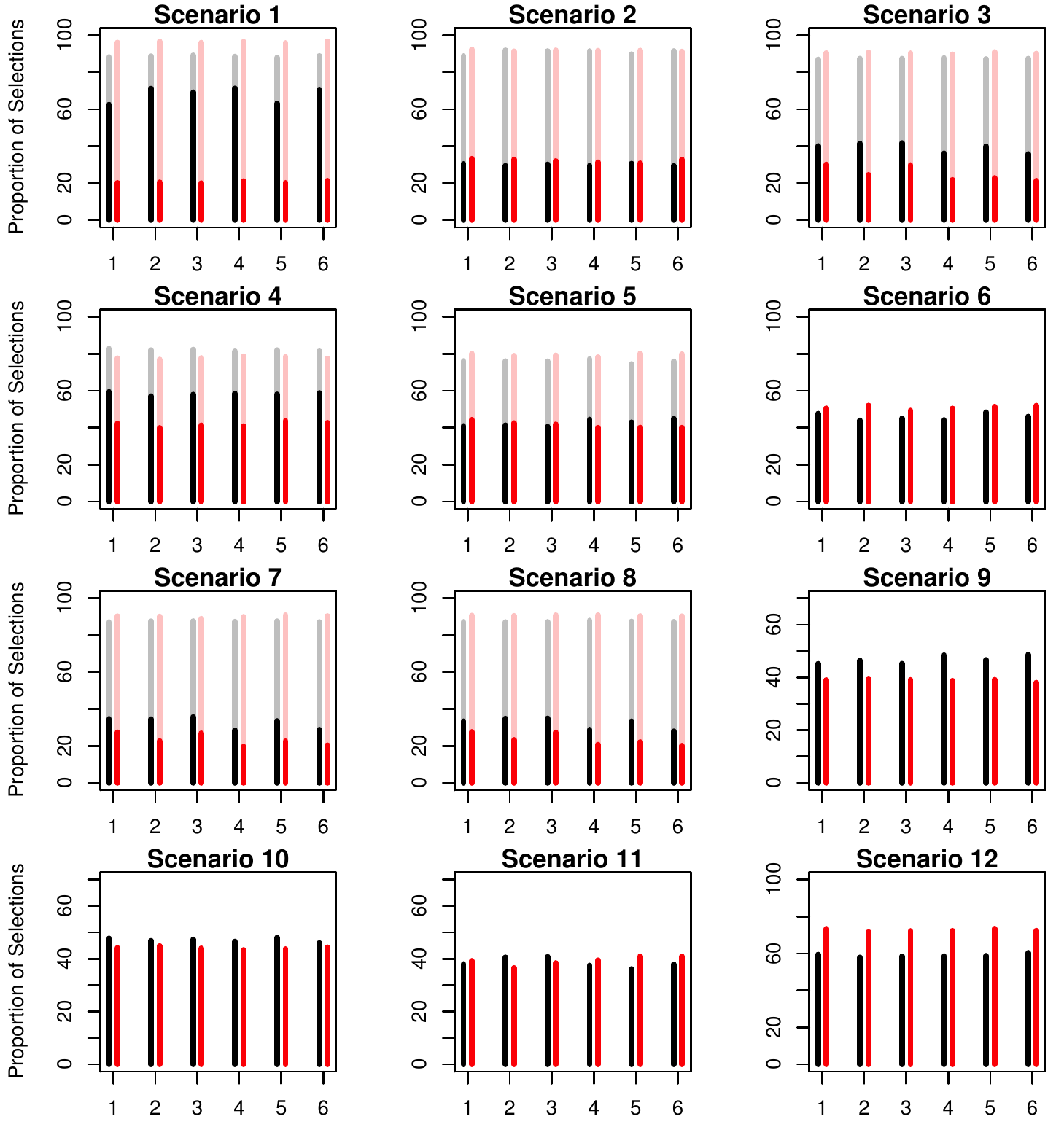}
 \caption{Proportion of optimal (bright) and correct (transparent) recommendsations by WE(R) (black) and WT (red) designs in scenario 1-12 across six permutations. The results are based on $10^4$ replications. \label{fig:motivation_result}}
\end{figure}

The results of the comparison in scenarios 1-12 using $6$ permutations are summarized in Figure \ref{fig:motivation_result}.
The length of the bar corresponds to the proportion of the optimal (solid) and correct (transparent) regimen recommendations by WE(R) (black) and WT (red). Overall, both designs are robust to toxicity ordering permutations under all scenarios. The difference of minimum and maximum proportions of the optimal and correct recommendations within one scenario does not exceed 5\% and 8\% for both designs, respectively. Regarding the proportion of \textit{correct} recommendations, the designs perform comparably (no more than 5\% difference) in scenarios with several correct regimens (scenarios 2-5, 7 and 8) with scenario 1 being an exception (8\% difference in favour to WT design). Regardless the large number of orderings, WT design preserves it accuracy in terms of the correct regimen recommendation.

Comparing \textit{optimal} regimen recommendations, WE(R) favours the optimal regimen over other correct regimens systematically. It results in superior characteristics in half of all scenarios: 1, 3, 4, 7-9 with the minimum difference across permutations ranging from 7\% (scenario 7) up to 45\% (scenario 1) and the maximum difference varying between 11\% and 52\%, respectively. Generally, WT is less conservative as it favours safe regimens with higher toxicity that results in a lower proportion of optimal recommendations in these scenarios. Both designs perform comparably in scenarios 2, 5,  10 and 11 within the maximum difference of 2-4\%. At the same time, less conservative nature of WT allows to outperform WE(R) by 3-8\% under scenario 6 in which the optimal regimen is the highest safe one. WT also shows a better performance in scenario 12 with difference in the proportion of optimal recommendations ranging in 10-13\%. The regimen 1 is optimal and  $36$ patients is not enough for WE(R) to investigate all regimens and come back to $T_1$, but it is enough for the WT to identify the decreasing ordering. 

Regarding scenarios 13 and 14 with no correct and no optimal regimens due to futility and high toxicity, WE(R) terminates the trial earlier in nearly 72\% and 85\% of the trials against 68\% and 79\%, respectively. WE(R) results in 10.5 and 10.3 toxic responses on average in scenarios 13 and 14 against 10.4 and 10.5 by WT, respectively. Regarding the efficacy outcomes, WE(R) results in 7.0 and 8.7 responses versus 7.4 and 9.0 by WT. This follows that both designs are able to terminate the trial with high probability and prevent unethical patient allocations. Investigating further ethical aspects of the design, the average numbers of toxicity and efficacy responses across all permutations in the rest of scenarios are given in Table \ref{tab:numbers_motivation}.

\begin{table}
   \caption{Mean number of toxicity and efficacy responses in scenarios $1-12$ across six permutations using $N=36$ patients and $M=6$ regimens. Figures corresponding to the best performance are in \textbf{bold}. The results are based on $10^4$ replicated trials. \label{tab:numbers_motivation}}
  \fbox{%
\begin{tabular}{ccccccccccccc}
 Scenario  & 1 & 2 & 3 & 4 & 5 & 6 & 7 & 8 & 9 & 10 & 11 & 12 \\
     \multicolumn{13}{@{}c}{{Toxicity responses}}\\
     WE(R) & \textbf{2.5} & 6.4 & \textbf{3.2} & \textbf{4.4} & \textbf{7.0} & \textbf{7.7} & \textbf{5.1} & \textbf{5.1} & \textbf{3.9} & 5.9 & \textbf{7.8} & 2.4 \\ 
  WT & 4.1 & \textbf{5.0} & 4.5 & 7.1 & 7.9 & 8.7 & 5.9 & 6.0 & \textbf{3.3} & \textbf{4.2} & \textbf{7.5} & \textbf{1.5} \\ 
    \hline
      \multicolumn{13}{@{}c}{{Efficacy responses}}\\
  WE(R) & 19.8 & \textbf{14.4} & \textbf{20.8} & 19.5 & \textbf{18.2} & 12.5 & \textbf{22.8} & \textbf{22.8} & \textbf{15.4} & \textbf{13.7} & \textbf{16.7} & 18.1 \\ 
  WT & \textbf{24.5} & \textbf{14.4} & \textbf{21.0} & \textbf{21.4} & \textbf{19.0} & \textbf{13.8} & \textbf{23.4} & \textbf{23.5} & \textbf{15.8} & \textbf{14.4} & \textbf{16.7} & \textbf{21.5} \\ 
 \end{tabular}}
\end{table}

The WE(R) design is generally more safe and results in at least nearly one less toxic response in scenarios 1, 3-8 with the largest difference of $2.7$ in scenario 4. WT design results in fewer toxicities in scenarios where the target regimen is among the first ones (scenarios 2, 10, 12). While the prior toxicity vector chosen for WE(R) suggests to proceed escalation, the model-based approach is able to identify the correct ordering faster. The average number of efficacy outcomes does not differ by more than one in the majority of scenarios (2, 3, 5, 7-11). Due to fact the WT waits for the complete efficacy response and due to model-based nature,  it results in 1-5 more average efficacy responses in scenarios 1, 4, 6 and 12. At the same time, WE(R) waits for the efficacy outcome and escalates slower as it was shown in Section \ref{sec:illustration}.

Summarizing, the proposed design is robust to the possible true toxicity and efficacy orderings. WE(R) is found to be a good and comparative alternative to the model-based design when large number of ordering are to be considered and the randomization might be hard to justify in an application. While all possible toxicity and efficacy orderings are still feasible to specify, it can be challenging to convince a clinician to randomize between them given the limited sample size of $N=36$ patients.  WE(R) is able to identify optimal and correct regimens with high probability while being more safe than model-based design and has comparable number of efficacies in majority of realistic scenarios. 

\subsection{Sensitivity analysis}
In the simulation study above, the toxicity and efficacy outcomes were assumed to be uncorrelated which may not hold in an actual trial. In this section, we investigate the robustness of the WE(R) design to the correlation in toxicity and efficacy under scenarios given in Figure \ref{fig:scenarios1}. The study was also conducted for different toxicity orderings permutations and the same qualitative results were obtained (not shown).

We follow the procedure proposed by \cite{tate1955} to generate correlated binary toxicity and efficacy outcomes. The procedure generates a binary normal vector with unit variances and pre-specified correlation coefficient $\rho$. The generated random variable is then transformed to a binary response by applying the cumulative distribution function and a quantile transformation, subsequently.
 
The results of the WE(R) performance in the case of high negative correlation ($\rho=-0.8$), an absence of correlation ($\rho=0.0$) and high positive correlation ($\rho=0.8$) under scenarios 1-12 are summarised in Table \ref{tab:corr}.
\begin{table}
   \caption{Operating charsteristics of the WE (R) design in scenarios 1-12: proportion of optimal and correct recommenations for different values of toxicity and efficacy outcomes correlation $r=\{-0.8,0.0,0.8\}$. The largest deviations are in \textbf{bold}. Results are based on $10^4$ replications. \label{tab:corr} }
   \footnotesize
  \fbox{%
  \begin{tabular}{c|cccccccccccccc}
 Scenario & 1 & 2 & 3 & 4 & 5 & 6 & 7 & 8 & 9 & 10 & 11 &12 \\
 \hline
     WE (R) &   \multicolumn{12}{@{}c}{{Proportion of optimal recommendations}}\\
$\rho=-0.8$ & 70.7 & \textbf{22.5} &  37.7 & 59.8 & 40.0 & 52.8 & 36.2 & \textbf{24.7} & {51.1} & 45.9 & {45.1} & 56.0  \\
$\rho=0.0$ & 66.5 & 31.0 & 39.5 & 59.3 & 40.5 & 46.8 & 33.2 & 32.8 &  44.8& 47.7 &   38.5 & 59.1 \\
$\rho=0.8$ & 65.5 & \textbf{41.2} & 42.4 & 61.4 & 42.7 &  47.5 & 33.5 & \textbf{38.3} & 43.6 & 51.4 & 37.3 & 63.6\\
    \hline
    &   \multicolumn{12}{@{}c}{{Proportion of correct recommendations}}\\
$\rho=-0.8$ & 90.9 & \textbf{81.2} & 91.4 & 91.8 & 78.0 & 52.8 & 91.6  & 75.4  & 51.1 & 45.9 & 48.1 &56.0 \\
$\rho=0.0$ & 88.1 &  89.7 & 87.2 & 86.7 &  73.2 & 46.8 & 86.7 & 77.5 & 44.8 &  47.7 &38.5 & 59.1  \\
$\rho=0.8$ & 88.4 &  \textbf{96.6} & 84.8 & 84.5 & 73.6 & 47.5 & 86.2  & 80.9 & 43.6 &  51.4 &  37.3 & 63.6 \\
      \end{tabular}}
\end{table}
In the majority of scenarios, WE(R) is robust to both negative and positive correlations. The differences between the proportion of optimal recommendations in positively and negatively correlated cases do not exceed 8\% in 10 out of 12 scenarios with scenario 2 and 8 being exceptions. Moreover, the proportions of optimal recommendations in correlated cases never differ by more than 10\% compared to the uncorrelated case. A noticeable difference in the performance can be seen in plateau scenarios 2 and 8 in which the optimal regimens are among low regimens ($T_1$ and $T_2$, respectively). In these scenarios efficacy probabilities are nearly the same. The negative correlation biases the recommendations to higher regimens which worsen the proportion of optimal recommendations by nearly 9\% in scenarios 2 and 8. In contrast, the positive correlation biases selections to lower regimens and leads to an increase by 10\% and 6\% in the proportion of optimal recommendations in these scenarios. There are no noticeable changes in the rest of scenarios as the bias caused by the correlation is smaller than the difference in true toxicity and probability estimates. 


Overall, the proposed design is robust to the highly correlated toxicity and efficacy outcomes. The proportion of optimal and correct recommendations in the correlated cases never differs by more than 10\% compared to the uncorrelated case. Despite assumed independence in the estimates, WE(R) is able to find the optimal and correct regimens in highly correlated cases. \\

 \section{Conclusions}
In this article, we have introduced a novel phase I/II design for molecularly targeted agents that does not require an assumption of monotonicity. The proposed design is based on the simple and intuitively clear trade-off function which can be easily computed by non-statisticians. The simulation results demonstrate that the novel approach can identify the optimal regimen with high probability and leads to ethical patient allocations. Therefore, it can be recommended for the clinical application with a limited sample size and missing, delayed efficacy responses. The novel design is applied to the challenging combination-schedule trial with uncertainty in both toxicity and efficacy ordering.  The proposal is shown to be a good alternative to the model-based designs when the ordering specification is challenging.

In this paper we have considered the setting where efficacy can only be observed if no toxicity is observed in a patient. At the same time, a setting with four possible outcomes can be also considered using the same information-theoretic approach. The only difference is the form of the Dirichlet distribution (\ref{pdf}) which is used to obtain the trade-off function \citep{entropy}. Additionally, the influence of various techniques of the missing and delayed efficacy outcomes implementation can be of interest in a setting with no parametric model. While binary outcomes are considered only, a continuous efficacy endpoint becomes more common choice in a clinical practice. An extension of the proposed design for a this case is the subject to a future research.

\section*{Acknowledgement}

The authors acknowledge the insightful and constructive comments made by associate editor and two reviewers. These comments have greatly helped us to sharpen the original submission. This project has received funding from the European Union’s
Horizon 2020 research and innovation programme under the 
Marie Sklodowska-Curie grant agreement No 633567 and by Prof Jaki's Senior Research Fellowship (NIHR-SRF-2015-08-001) 
supported by the National Institute for Health Research. The views expressed in this publication are those of the authors and not necessarily those of the NHS, the National Institute for Health Research or the Department of Health.

\section*{Supplementary Materials}

Supplementary Materials provide details on the calibration of design parameters and on the performance of the WE design in the context of a single-agent trial including the case of efficacy data available earlier.

\bibliographystyle{rss}
\bibliography{mybib}

\clearpage 
\section*{Supplementary Materials}

\section{Application of WE design to a single-agent trial}
The proposed design can be applied to a wide range of Phase I/II clinical trials. While the performance of the WE design is demonstrated in the context of the motivating trial, it can be also applied to a single agent dose-finding trial for which several model-based designs were recently proposed \cite[see e.g.][]{wages2015,riviere2016}. Here we show the comparison of the proposed design to the currently used and provide a step-by-step algorithm how the parameters of the proposed design can be calibrated.
\subsection{Simulation setting}
We consider $M=6$ doses and $N=60$ patients. The dose-toxicity relationship is known to be a non-decreasing function, but a clinician expects either a plateau or an umbrella shape for the dose-efficacy curve. A toxicity is evaluated after three weeks while an efficacy outcome is evaluated after six weeks. To conduct the trial in a timely manner, the next cohort of patients is allocated after the toxicity data for previous cohort are available. The upper toxicity and the lowest efficacy bounds are $\phi=0.35$ and $\psi=0.20$. The goal is to study the ability of the WE design to identify the \textit{optimal} and \textit{correct} doses. A dose is called \textit{optimal} if it is safe, has maximal efficacy and minimal toxicity while a safe dose with maximum efficacy (irrespective of it also having lowest toxicity) is called \textit{correct}. 

We consider 14 scenarios that were used for the motivating trial simulations: eight plateau scenarios (1-8) suggested by \cite{riviere2016}, $4$ umbrella shaped scenarios (9-12) studied in \cite{wages2015} and two scenarios with no correct doses (13-14, due to inefficacy and toxicity, respectively)  $-$ see Figure 4 in the main paper.

In the analysis we focus on (i) the proportion of optimal/correct recommendations, (ii) the average number of toxic responses, (iii) the average number of efficacy responses. The study is performed using \texttt{R} \citep{Rcore} and 10,000 replications for each scenario. We compare the characteristics with the `MTA` design proposed by \cite{riviere2016} and the `WT` design developed by \cite{wages2015}. Parameters of the designs are chosen as in the original proposals with an exception of using cohort size $c=3$  and $80\%$ confidence intervals for stopping rules for the WT design.

\subsection{Design specification}
As before, we use a target toxicity of $\gamma_t=0.01$ and a target efficacy of $\gamma_e=0.99$. Due to the known toxicity ordering, the design is restricted to satisfy the coherence principals with $q=1$. While we consider both non-randomized and randomized versions of the WE design to study an allocation rule impact, the design specification for the  non-randomized WE design is provided only.

\subsubsection{Prior}
Parameters $\beta_{t,i}=\beta_{e,i}=1$ of the prior Beta distribution in (8) are chosen for all dose levels $i=1,\ldots,M$ to emphasize a limited available information. Parameters $\nu_{t,i}$ and $\nu_{e,i}$ (which coincide the prior probabilities of toxicity and efficacy for $\beta_{t,i}=\beta_{e,i}=1$) are specified such that the WE design leads to accurate optimal dose recommendation in various different scenarios. The prior values of $\nu_{t,i}$ and $\nu_{e,i}$ are calibrated over scenarios 1-8 with different locations of the optimal and correct doses. There are two restrictions on the prior parameters: the escalation should start at the first dose and no dose skipping is allowed. To restrict number of possible parameters to be calibrated over, we assume that prior efficacy and toxicity probability increases linearly as $\nu_{t,i}=start_t+w_t \times i$ and  $\nu_{e,i}=start_e+w_e \times i$. Then, we search for the values of $start_t,start_e, w_t, w_e$ such that the geometric mean of the proportion of optimal selection over all scenarios is maximised. 

Prior vectors of toxicity probabilities $\hat{\bp}_t^{(0)} = \left[0.05,0.14,0.23,0.32,0.41,0.50 \right]^{\rm T}$ and efficacy probabilities $\hat{\bp}_e^{(0)} = \left[0.55,0.58,0.61,0.64,0.67,0.70 \right]^{\rm T}$ are subsequently used for the non-randomized WE design.

Similarly, vectors of prior toxicity $$\hat{\bp}_t^{(0)} = \left[0.25,0.35,0.45,0.55,0.65,0.75 \right]^{\rm T}$$  and $$\hat{\bp}_e^{(0)} =  \left[0.65,0.69,0.73,0.77,0.81,0.85\right]^{\rm T}$$ efficacy  probabilities are used for the randomized WE(R) design. It was found that the randomised WE(R) design is more robust to the choice of the prior parameter than non-randomised WE.

\subsubsection{Safety constraint}
\begin{figure}[!ht] 
\centering
\includegraphics[width=1\textwidth]{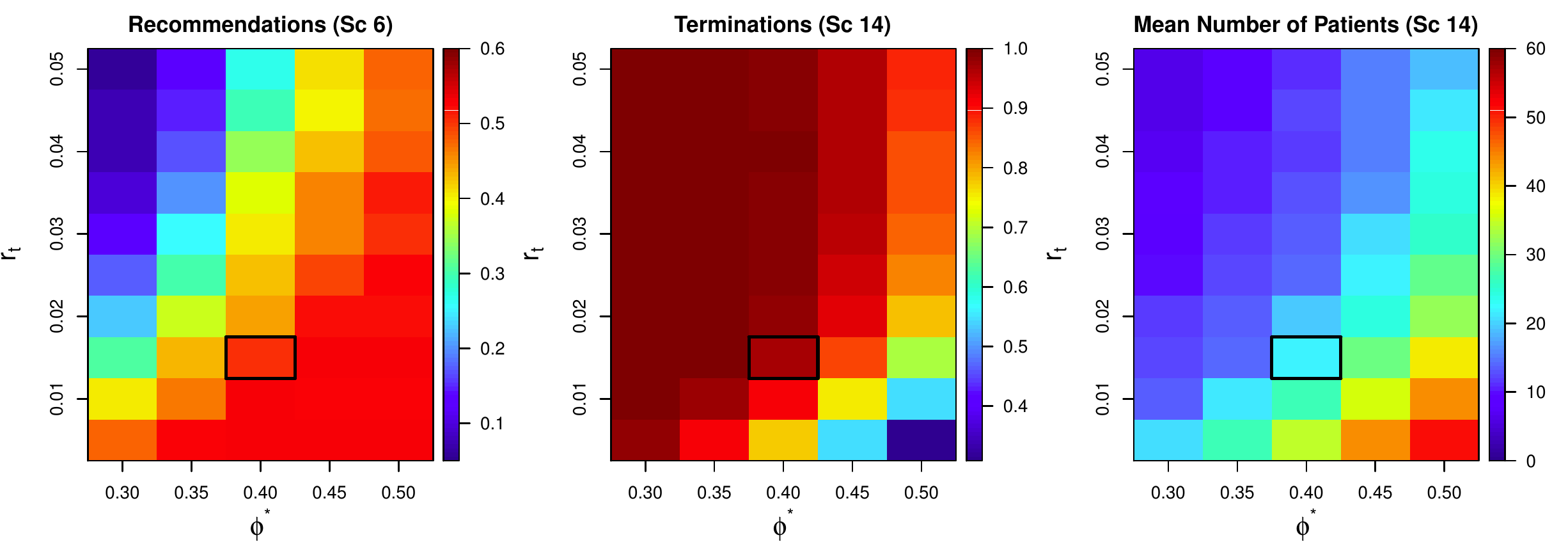}
 \caption{Safety constraint parameters calibration: $\phi^* \in (0.3,0.5)$,  $r_t \in (0,0.05)$ in scenarios 6 and 14. The proportion of correct recommendations (terminations) and the mean number of patients in a trial (scenario 14). The final choice is marked by a black frame. Results are based on $10^4$ replications. \label{fig:safety}}
\end{figure}

To set the time-varying safety constraint, we use $\zeta_{N}=0.30$ and calibrate $\phi^*$, $r_t$ using the highly toxic scenario $14$ and the flat scenario 6. These two scenarios are chosen to represent the trade-off in the safety constraint. The proportion of correct recommendations (terminations) and mean number of patients involved in a trial for different parameters values are given in Figure \ref{fig:safety}.
The mean number of patients in scenario 6 does not vary a lot and the corresponding graph is not shown. In scenario 6 the highest proportion of the optimal recommendations corresponds to the least strict safety constraint (right bottom corner), but only  35\% of trials in scenario 14 are then terminated. At the same time, the most strict rule (left top corner) results in 100\% of terminations in scenario 14, but only in $5\%$ of correct recommendation in scenario 6. Parameters $r_t=0.0125$ and $\phi^*=0.4$ are chosen for subsequent study as a reasonable trade-off. The same parameters of the safety constraints are used for the randomized design.

\subsubsection{Futility constraint}
\begin{figure}[!ht] 
\centering
\includegraphics[width=1\textwidth]{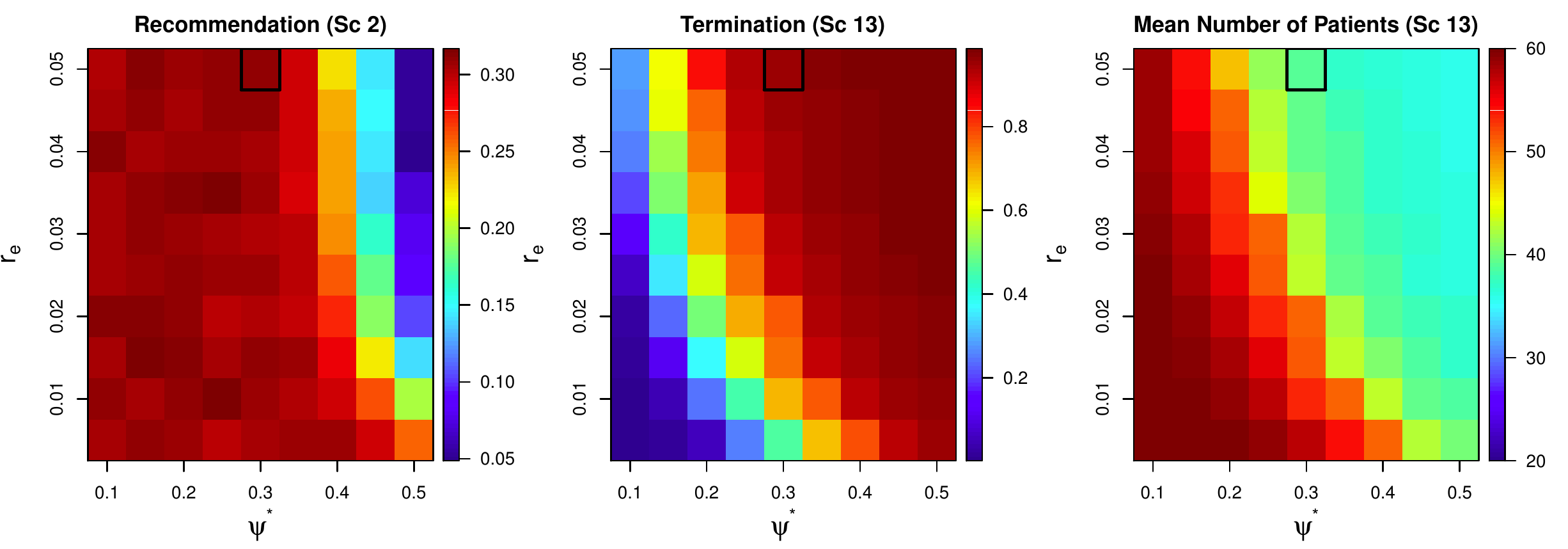}
 \caption{Futility constraint parameters calibration: $\psi^* \in (0.1,0.5)$ and $r_e \in (0,0.05)$ in scenarios 2 and 13. The proportion of correct recommendations (terminations) and the mean number of patients in a trial (scenario 13). The final choice is in the black frame. Results are based on $10^4$ replications. \label{fig:futility}}
\end{figure}

We calibrate the futility constraint by fixing $\xi_{N}=0.50$ and tuning $\psi^*$ and $r_e$ using two opposite scenarios - 2 and 13. In scenario 2 all doses have the same efficacy probability. In scenario 13 there are no correct doses as all efficacious doses have unacceptable toxicity. The proportion of correct recommendations (terminations) and the mean number of patients are given in Figure \ref{fig:futility}.
Since the mean number of patients in scenario 2 does not vary this graph is not shown. A stricter constraint is favourable in scenario 13 and less favourable in scenario 2 while the opposite is true for less strict constraints. Subsequently, parameters $\psi^* = 0.3$ and $r_e=0.05$ are used for both non-randomized and randomized designs.

\subsection{Operating characteristics}
The results of the comparison in scenarios 1-8 with plateau dose-efficacy relation and in scenarios 13-14 with no correct doses are summarized in Table \ref{tab:result1} and Table \ref{tab:numbers}. Each figure in Table \ref{tab:result1} corresponds to proportions of optimal or correct dose recommendations. The detailed results, such as the selection proportions and mean number of patients on each dose, are given in Table \ref{tab:app1} and Table \ref{tab:app3} in Appendix.

\begin{table}
   \caption{ \label{tab:result1}  Proportion of optimal and correct dose recommendations in scenarios 1-8 and 13-14  using $N=60$ patients and $M=6$ doses for all considered designs. Figures corresponding to the highest performance in each scenario are in \textbf{bold}. The results are based on $10^4$ replicated trials.}
  \fbox{%
\begin{tabular}{ccccccccccc}
Scenario  & 1 & 2 & 3 & 4 & 5 & 6 & 7 & 8 & 13 & 14 \\
 \hline
      \multicolumn{11}{@{}c}{{Proportion of the optimal dose recommendation}}\\
       ${\rm WE}$ & {58.8} & 30.3 & \textbf{65.8} & {\textbf{71.1}} & {\textbf{60.3}}  & 53.5 & {\textbf{60.0}}  & {37.8} & {\textbf{95.2}} & {\textbf{96.9}} \\
${\rm WE \ (R)} $ & {\textbf{72.0}} & 35.0 & 51.0 & {\textbf{69.9}} & {54.5} & 56.7 & {48.5}  & {36.4} & {\textbf{93.2}} & {\textbf{97.4}} \\
      ${\rm MTA}$ & {57.0} & {\textbf{60.2}} & 48.4 & 53.7 & {55.3} & 55.9 & 37.9 & \textbf{43.0 } & 91.9 &  91.0 \\
       ${\rm WT}$ & 19.6 & {41.9} & 29.3 & 25.3 & 27.0 & {\textbf{65.2}} & 27.1 & 26.1 & 91.5 & 90.4 \\
    \hline
      \multicolumn{11}{@{}c}{{Proportion of the correct dose recommendation}}\\
       ${\rm WE}$ & 60.3 & 90.0 & 87.5 &79.8  & {\textbf{89.7}} & 53.5 & 77.8 & {\textbf{91.5}} & - & - \\
${\rm WE \ (R)} $ & 89.3 &{\textbf{97.0}} & \textbf{92.8} & {\textbf{89.1}} & {\textbf{87.5}} & 56.7 & {88.3}  & {\textbf{91.1}} & - & - \\
      ${\rm MTA}$ & {94.1} & {\textbf{96.6}} & 83.8 & 82.3 & 80.6 &55.9 & {89.9} & 77.4 & - & - \\
       ${\rm WT}$ & {\textbf{98.1}} & {\textbf{97.6}} & \textbf{93.6} & {86.5} & 80.0 &{\textbf{65.2}} & {\textbf{93.3}} & 81.1 & - &  -\\
      \end{tabular}}
\end{table}

With respect to the \textit{optimal} dose recommendation, both versions of the proposed design perform comparably or better than model-based designs in the majority of scenarios. The WE design without randomization leads to a considerable improvement in scenarios 3-5, 7 and 13-14 and outperform the best model-based alternative by up to 20\%. While the randomized WE(R) shows the comparable to the best model-based alternative  performance in scenarios 3,5 and 6, it also results in more accurate optimal dose recommendations in scenarios 1, 4, 7, 13-14. However, both WE and WE(R) are outperformed by MTA in scenarios 2 and 8 in which dose-toxicity and dose-efficacy curves are flat in the neighbourhood of the optimal dose. While MTA recommends the lowest dose by default, such small differences in toxicity and efficacy probabilities are difficult to find the small sample size. At the same time, the absence of a parametric model is not found to be a problem in any other cases. WT design outperforms all other designs in scenario 6 with the optimal dose being the highest safe one. Generally, WT is less conservative as it favours safe doses with higher toxicities that results in a low proportion of optimal recommendations if the optimal dose is not the highest safe one (see also Table \ref{tab:app1} and Table \ref{tab:app3}).

Considering the proportion of the \textit{correct} recommendations, WT outperforms MTA in all scenarios and has the best performance among all alternatives in scenarios 1-3 and 6-7. In the rest of scenarios WT has either comparable or worse performance than the randomized WE(R). Comparing WE and WE(R), the randomized design is more robust in the correct recommendations with a largest difference in scenario 1. Here, the chosen prior would not escalate to dose $6$ once the optimal is already find at dose $5$ if no randomization is used.

In terms of toxicities we find that the non-randomized WE design results in considerably lower number of toxicities in almost all scenarios with the largest difference observed in scenario 4.  As the WT approach is less conservative, it results in a greater number of toxicities, but also leads to the highest average number of efficacies in all scenarios. In contrast, the cost of the WE's lowest number of toxicities is a smaller number of efficacies. In scenarios 13 and 14 with no optimal and correct doses all alternatives result in nearly the same average number of toxicities and efficacies.
 
 \begin{table}
   \caption{Mean number of toxicity and efficacy responses in scenarios $1-8$ and $13-14$ using $N=60$ patients and $M=6$ doses for all considered designs. The results are based on $10^4$ replicated trials. \label{tab:numbers}}
  \fbox{%
\begin{tabular}{ccccccccccc}
Scenario   & 1 & 2 & 3 & 4 & 5 & 6 & 7 & 8 & 13 & 14 \\
   \hline
     \multicolumn{11}{@{}c}{{Toxicity responses}}\\
       ${\rm WE}$ & \textbf{3.1} & \textbf{6.0} & \textbf{2.8} & \textbf{4.5} & \textbf{7.9} & \textbf{11.0} & \textbf{5.9} & \textbf{5.6} & \textbf{11.7} & \textbf{11.0} \\
 ${\rm WE \ (R)} $ & 4.0 & 6.9 & 4.3 & 6.0 & 8.7 & \textbf{10.8} & 6.9 & 6.9 & 13.0 & \textbf{10.9} \\
       ${\rm MTA}$ & 5.5 & 8.1 & 6.0 & 10.0 & 12.1 & 13.2 & 9.6 & 9.3 & \textbf{11.0} & \textbf{11.5} \\
        ${\rm WT}$ & 6.8 & 6.7 & 7.3 & 13.2  & 13.5 &14.7 & 10.0 & 9.1 & \textbf{11.2} & 12.1 \\
      \hline
       \multicolumn{11}{@{}c}{{Efficacy responses}}\\
      ${\rm WE}$ & 28.5 & \textbf{24.0} & 33.0 & 30.0  & 29.8 & 19.4 & 34.6 & \textbf{28.9} & 6.0 & \textbf{9.1} \\
${\rm WE \ (R)}$ & 27.4 & \textbf{24.0} & \textbf{34.5} & 32.2 & 29.8 & 19.2 & 36.5 & \textbf{29.2} & \textbf{7.1} & \textbf{8.9} \\
     ${\rm MTA}$ & 38.0 & \textbf{24.0} & \textbf{34.6} & 35.0 & 29.4 &21.4 & \textbf{39.1} & \textbf{29.3} & 6.2 & \textbf{9.6} \\
      ${\rm WT}$ & \textbf{41.5} & \textbf{24.0} & \textbf{35.5} & \textbf{37.0} & \textbf{32.3} & \textbf{24.4} & \textbf{39.9} & \textbf{29.2} & 5.4 & \textbf{9.7} \\
      \end{tabular}}
\end{table}

The results of the comparison in scenarios 9-12 with an umbrella shaped dose-efficacy relationship and only one correct dose are given in Table \ref{tab:result2}. Overall, WE designs have more robust optimal dose identification in non-monotonic scenarios. The WE design with no randomization outperforms MTA by up to 35\% and WT by up to 6\%. WT has the highest proportion of the optimal dose recommendations in scenarios 11 with nearly 10\% difference with the non-randomized WE. The MTA design is more conservative and recommends $d_1$ with the highest probability that results in the best performance in scenario 12, but poor performance in other cases. The non-randomized WE is favourable compared to the randomized version due to the single correct dose in each scenario. The average number of toxicities of the WE design is again the safest alternative. In contrast to the scenarios with plateau, it can now also result in a larger number of efficacy responses (e.g. in scenario 9) due to the non-monotonic shape of the dose-efficacy curve. 

\begin{table}
   \caption{ \label{tab:result2}  Proportion of optimal dose recommendations, mean number of toxicity and efficacy responses in scenarios 9-12  using $N=60$ patients and $M=6$ doses for all considered designs. The results are based on $10^4$ replicated trials.}
  \fbox{%
  \begin{tabular}{ccccc|cccc|cccc}
Scenario  & 9 & 10 & 11 & 12 & 9 & 10 & 11 & 12 & 9 & 10 & 11 & 12  \\
 \hline
     & \multicolumn{4}{@{}c}{{Optimal recommendation}} & \multicolumn{4}{@{}c}{{Toxicity responses}} & \multicolumn{4}{@{}c}{{Efficacy responses}} \\
       ${\rm WE}$ & \textbf{54.7} & \textbf{55.9} & 46.5 & 80.1 & \textbf{4.5} & \textbf{5.7} & \textbf{10.0} & \textbf{1.8} & \textbf{29.7} & \textbf{25.6} & 27.4& 35.4   \\
${\rm WE \ (R)} $ & \textbf{56.7} & \textbf{56.2} & 47.9 & 70.9 &5.5 & 6.7 & \textbf{10.0} & 3.3 & 28.4 & 24.6 & 27.1 & 32.7 \\
      ${\rm MTA}$ & 20.3 & 35.3 & 46.0 & \textbf{96.1} & 5.0 & 6.3 & 12.8 & 2.6 & 26.3 & 23.9 & 28.7 & 35.5 \\
       ${\rm WT}$ & 50.1 & 49.3  & \textbf{56.9}  & 75.9 & 5.5 & 5.9 & 12.2 & 2.4 & 27.9 & 24.8 & \textbf{29.4} & \textbf{36.3}\\
      \end{tabular}}
\end{table}

Overall, the proposed approaches have better or comparable operating characteristics in 9 out of 14 considered scenarios even with less information used in a trial. Comparing two assignment rule of the WE design, the non-randomized WE is always less accurate in terms of the correct dose identification. As the result, the WE design without randomization should be preferred if  only one correct dose is expected or a clinician is cautious about toxicity profile, while the randomized WE is a robust choice if multiple \textit{correct} doses are expected. 

\subsection{Early efficacy data}

In the setting above, it is assumed that it takes twice as long to observe the efficacy outcome than the toxicity endpoint. It is, however, possible that an efficacy (or lack of efficacy) can be observed at the time of the interim analysis for some of the patients. As the proposed design includes all available information, it can also accommodate earlier efficacy (no efficacy) data. This section we study how the operating characteristics of the non-randomised WE design are affected if a certain proportion of `no efficacy` responses can be observed earlier.

The setting above remains unchanged with the following exception: if the patient has observed no DLT and will have `no efficacy `, it is assumed that the outcome can be observed at the time of toxicity evaluation with probability $\pi$. If observed earlier, the WE design uses this information for the next patient allocation. We consider two cases: $\pi=0$ (the original setting) and $\pi=1/2$ (half of `no efficacies` can be seen earlier). The results are given in Table \ref{sensitive2}.

\begin{table}
   \caption{Operating characteristics of WE in scenarios 1-12 with no early efficacy data available ($\pi=0$) and with half `no efficacy`  outcomes ($\pi=1/2$) available at the time of toxicity evaluation: recommendation proportions, mean number of toxicity (T) and efficacy (E) responses . The optimal dose is in \textbf{bold} and correct doses are \underline{underlined}. Results are based on $10^4$ replications. \label{sensitive2} }
\centering
\begin{tabular}{rrrrrrrrr}
  \hline
WE & $d_1$ & $d_2$ & $d_3$ & $d_4$ & $d_5$ & $d_6$ & T & E \\ 
  \hline
    \multicolumn{8}{@{}c}{{Scenario 1}}\\
  & (.005;.01) & (.01;.10) & (.02;.30) & (.05;.50) & \underline{\textbf{(.10;.80)}} & \underline{{(.15;.80)}} \\
   $\pi=0$ & 0.0 & 0.1 & 2.3 &37.4& \textbf{58.8} & 1.5 &  3.1 &  28.5   \\
 $\pi=1/2$ & 0.0 & 0.0 & 1.5 & 23.5 & \textbf{64.9} & 10.0 & 4.3 & 34.8 \\ 
   \hline
       \multicolumn{8}{@{}c}{{Scenario 2}}\\
  & \underline{\textbf{(.01;.40)}} & \underline{(.04;.40)} & \underline{(.10;.40)} & \underline{(.25;.40)} & (.50;.40) & (.70;.40)\\
       $\pi=0$ &  \textbf{30.3} & 26.3 &  20.8 &  12.5  & 6.3 & 3.7  &6.0 & 24.0  \\
   $\pi=1/2$  & \textbf{33.2} & 27.1 & 21.9 & 12.2 & 3.7 & 1.6 & 7.2 & 24.0 \\ 
   \hline
        \multicolumn{8}{@{}c}{{Scenario 3}}\\
  & (.01;.25) & (.02;.45) & \underline{\textbf{(.05;.65)}} & \underline{(.10;.65)} & \underline{(.20;.65)} & \underline{(.30;.65)}\\
     $\pi=0$ &  0.6 & 12.6  & \textbf{65.8} & 18.2  & 2.8 &  0.1  & 2.8 & 33.0  \\
 $\pi=1/2$ & 0.9 & 9.4 & \textbf{57.7} & 24.0 & 6.8 & 1.1 & 3.7 & 34.8 \\ 
    \hline
     \multicolumn{8}{@{}c}{{Scenario 4}}\\
  & (.01;.05) & (.02;.25) & (.05;.45) & \underline{\textbf{(.10;.70)}} &\underline{(.25;.70)} & (.50;.70) \\
     $\pi=0$  &0.0 & 0.5 & 19.6 & \textbf{71.1}  & 8.5  & 0.3  &4.5 & 30.0  \\
 $\pi=1/2$ & 0.0 & 0.9 & 14.6 & \textbf{68.0} & 15.8 & 0.7 & 5.9 & 33.9 \\ 
     \hline
      \multicolumn{8}{@{}c}{{Scenario 5}}\\
 & (.01;.10) & (.05;.35) & \underline{\textbf{(.15;.60)}} & \underline{(.20;.60)} & (.45;.60) & (.60;.60)\\
     $\pi=0$&  0.1 & 6.2 & \textbf{60.3} & 28.9  & 3.6 &  0.8  & 7.9 &  29.8  \\  
  $\pi=1/2$ & 0.1 & 6.3 & \textbf{56.8} & 32.8 & 3.3 & 0.7 & 9.2 & 31.5 \\  
    \hline
              \multicolumn{8}{@{}c}{{Scenario 6}}\\
  & (.01;.05) & (.05;.10) & (.10;.20) & (.20;.35) & \underline{\textbf{(.30;.55)}} & (.50;.55)\\
    $\pi=0$  & 0.4 &  0.8 & 3.7 & 18.9 & \textbf{53.5} & 18.9  & 11.0 & 19.4   \\
  $\pi=1/2$ & 1.4 & 1.4 & 4.9 & 22.2 & \textbf{57.2} & 10.0 & 13.1 & {22.2} \\ 
    \hline
   \multicolumn{8}{@{}c}{{Scenario 7}}\\
 & (.02;.30)& {(.07;.50)}& \underline{\textbf{(.13;.70)}}& \underline{(.17;.73)} & \underline{(.25;.76)} & \underline{(.30;.77)} \\
  $\pi=0$  &  0.5 &  21.4 &\textbf{60.0} & 16.2 & 1.7   & 0.0 & 5.9 & 34.6 \\
  $\pi=1/2$  & 0.9 & 16.4 & \textbf{53.4} & 23.2 & 5.5 & 0.6 & 6.9 & {37.0} \\ 
      \hline
              \multicolumn{8}{@{}c}{{Scenario 8}}\\
 & (.03;.30) & \underline{\textbf{(.06;.50)}} & \underline{(.10;.52)} & \underline{(.20;.54)}& (.40;.55) & (.50;.55) \\
    $\pi=0$  &  3.2  & \textbf{37.8} &  34.7 & 19.1 &  4.2 & 1.0  & 5.6 & 28.9  \\
  $\pi=1/2$  & 3.0 & \textbf{37.8} & 34.0 & 20.1 & 4.0 & 1.1 & 6.9 & 29.5 \\ 
   \hline
               \multicolumn{8}{@{}c}{{Scenario 9}}\\
 & (.01;.30) & (.05;.50) & \underline{\textbf{(.10;.60)}} & (.15;.40) & (.20;.25) & (.25;.15)\\
  $\pi=0$ &   3.0 & 34.7  & \textbf{54.7} &  5.8  & 1.3   & 0.5 &4.4 & 29.7 \\
     $\pi=1/2$ & 3.8 & 34.2 & \textbf{54.6} & 6. & 1.1 & 0.2 & 5.1 & 29.7 \\ 

     \hline
         \multicolumn{8}{@{}c}{{Scenario 10}}\\
 & (.02;.38) & \underline{\textbf{(.06;.50)}} & (.12;.40) & (.30;.30) & (.40;.25) & (.50;.20)\\
    $\pi=0$  &  18.4 & \textbf{55.9}  &15.9 &  3.9  & 3.2   & 2.7 &5.7 & 25.6 \\
      $\pi=1/2$ & 20.2 & \textbf{57.1} & 17.2 & 2.8 & 1.6 & 1.0 & 6.3 & 25.5 \\ 

     \hline
             \multicolumn{8}{@{}c}{{Scenario 11}}\\
         & (.03;.25) & (.09;.35) & (.16;.48) &  \underline{\textbf{(.28;.65)}} & (.42;.52) & (.56;.39) \\
        $\pi=0$  & 2.2 & 9.8  &30.4 & \textbf{46.5}  & 8.6 & 2.8  &10.0 & 27.4 \\
         $\pi=1/2$ & 3.5 & 10.8 & 31.7 & \textbf{45.5} & 6.3 & 1.8 & 11.4 & 28.8 \\ 

                \hline
              \multicolumn{8}{@{}c}{{Scenario 12}}\\
         & \underline{\textbf{(.02;.68)}}& (.05;.56) & (.07;.49) & (.09;.40) & (.11;.33) & (.13;.26) \\
   $\pi=0$  &   80.1 &  14.5 &  3.9 & 1.1  & 0.3  & 0.1  & 1.8 & 35.4  \\
       $\pi=1/2$ & \textbf{74.1} & 17.5 & 5.8 & 1.8 & 0.6 & 0.1 & 2.2 & 36.8 \\ 

\end{tabular}
\end{table}

As expected, the availability of some of the efficacy information earlier leads to a less conservative design that allows more rapid escalation. Earlier `no efficacy` data even in half of the patients lead to more ethical patient allocation. This can be seen by increased numbers of efficacies almost in all scenarios with the cost of reasonable increase in the average number of toxicity responses.  The largest increase can be seen in scenario 1 where the average number of efficacy response increase by nearly 7, while toxicity increases only by 1. The information about earlier efficacy also improves the proportion of optimal recommendations in the scenarios where the target dose is high - by 6\% in scenario 1 and by 4\% in scenario 6. As the design being less conservative it favours higher doses among correct ones. This decreases the proportion of optimal recommendations in scenario 3, 5, 7 and 12 by 3-7\%. At the same time, the proportion of correct recommendations is either unchanged (scenario 5 and 8) or increased by at least 5\% (all the rest plateau scenarios). This confirms that the WE design in the setting with no earlier efficacy information is more conservative, but the difference in correct selection is relatively small.

\clearpage

\section*{Appendix}

\begin{table}
   \caption{Operating charsteristics of WE, WE(R), MTA and WT design in scenarios 1-5: recommendation proportions, mean number of patients assigned to a dose (in brackets), termination proportion (Term), mean number of toxicity (T) and efficacy (E) respones. The optimal dose is in \textbf{bold} and correct doses are \underline{underlined}. Results are based on $10^4$ replications. \label{tab:app1} }
       \footnotesize
  \fbox{%
\begin{tabular}{cccccccccc}
   & $d_1$ &   $d_2$ &  $d_3$ &   $d_4$ &   $d_5$ & $d_6$ & Term & T & E\\
   \hline
    \multicolumn{8}{@{}c}{{Scenario 1}}\\
  & (.005;.01) & (.01;.10) & (.02;.30) & (.05;.50) & \underline{\textbf{(.10;.80)}} & \underline{{(.15;.80)}} \\
    ${\rm WE}$ & 0.0 & 0.1 & 2.3 &37.4& 58.8 & 1.5 & 0.0 &  3.1 &  28.5   \\
   & (6.1) & (6.3) &  (9.5)  & (20.5) & (17.4) &  (0.3) \\
      ${\rm WE (R)}$ & 0.0 & 0.2 & 1.0  & 9.5 & 72.0 & 17.3 & 0.0 & 4.0 & 27.4  \\
     &  (5.1) & (5.0) & (8.2) & (13.8)  & (21.7)  &  (6.2) \\
   ${\rm MTA}$ &  0.0 & 0.0  & 0.7  & 4.5 &  57.0  & 37.1 &0.7 & 5.5 & 38.0 \\
  & (3.3) &  (3.7)  & (4.8)  & (7.7) &  (21.5) &  (19.0) \\
           ${\rm WT}$ & 0.0	& 0.1 &	0.4	& 1.3	& 19.6	& 78.5  & 0.1 & 6.8 & 41.5 \\
          & (3.9)	 & (1.6)	 & (2.4)	& (3.6)	 & (11.4)	 & (37.1)	 \\
    \hline
        \multicolumn{8}{@{}c}{{Scenario 2}}\\
  & \underline{\textbf{(.01;.40)}} & \underline{(.04;.40)} & \underline{(.10;.40)} & \underline{(.25;.40)} & (.50;.40) & (.70;.40)\\
       ${\rm WE}$ &  30.3 & 26.3 &  20.8 &  12.5  & 6.3 & 3.7 & 0.2 &6.0 & 24.0  \\
        & (18.3)  & (17.0) &(13.7) & (7.9) & (2.7) & (0.4) \\
             ${\rm WE (R)}$ & 35.0 & 29.0 & 21.5 & 11.5  & 2.8 &  0.2 &0.1 & 6.9 & 24.0  \\
         &    (15.0)  & (15.7) &  (15.7) &  (10.0) &   (3.3)  &  (0.4) \\
   ${\rm MTA}$ &  60.2 &  20.3  & 8.8  & 7.3 & 2.6 & 0.3 & 0.6 & 8.1 & 24.0  \\
   & (18.8) & (13.0)  & (10.7) & (10.7) & (5.8) &  (0.8) \\
           ${\rm WT}$ & 41.9 &	24.5	 &16.9 &	14.2	 &2.4	& 0.0 & 0.0 & 6.7 & 24.0	   \\
          & (23.1)	& (13.0)	& (10.7) &	 (8.6) &	(3.4) & 	(1.3) \\
      \hline
        \multicolumn{8}{@{}c}{{Scenario 3}}\\
  & (.01;.25) & (.02;.45) & \underline{\textbf{(.05;.65)}} & \underline{(.10;.65)} & \underline{(.20;.65)} & \underline{(.30;.65)}\\
    ${\rm WE}$ &  0.6 & 12.6  & 65.8 & 18.2  & 2.8 &  0.1 & 0.0 & 2.8 & 33.0  \\
    & (7.2)  & (15.9) &  (29.2) &   (6.7) &  (0.9) & (0.0) \\
          ${\rm WE (R)}$ & 0.7 & 6.6 & 51.0 & 30.5 & 10.6 & 1.2 &0.0 &4.3 & 34.5  \\
          & (6.4)  & (9.6) & (21.3)  &(16.1) & (5.8) &  (0.9) \\
   ${\rm MTA}$ & 2.0 &  14.3 &  48.4 &  19.2 &   9.8 & 6.4 & 0.0 & 6.0 &  34.6 \\
    & (6.3) &   (9.8)  & (15.5) &  (12.9)  & (10.4) & (5.1) \\
           ${\rm WT}$ & 1.4	& 4.9&	29.3	 &29.9 &	22.6	 &11.8	 & 0.0 & 7.3 & 35.5 \\
          &  (6.1)	 & (5.0)	& (13.8)& (14.8) &	(12.0)	 & (8.3)	 \\
          \hline
        \multicolumn{8}{@{}c}{{Scenario 4}}\\
  & (.01;.05) & (.02;.25) & (.05;.45) & \underline{\textbf{(.10;.70)}} &\underline{(.25;.70)} & (.50;.70) \\
    ${\rm WE}$ &0.0 & 0.5 & 19.6 & 71.1  & 8.5  & 0.3 &0.0 &4.5 & 30.0  \\
 &   (6.2) & (7.9)  &(17.8) &(25.2)&  (2.9)  &(0.1) \\
           ${\rm WE (R)}$ &  0.0  & 1.3  & 9.2  & 69.9 & 18.5  & 1.0 & 0.0 & 6.0 &  32.2 \\
          &  (5.4) &   (6.9) & (13.4) & (24.1) &  (9.2)  & (0.8) \\
   ${\rm MTA}$ & 0.0 & 0.7 &  8.2  & 53.7  & 28.6 &  8.5 & 0.4 & 10.0 & 35.0 \\
 &   (3.8)  & (5.0) & (9.1) &  (19.0) &  (15.9) & (7.0) \\
           ${\rm WT}$ &  0.0 &	0.4 & 	2.1 & 	25.3	 &61.2 & 	10.8	 & 0.2 & 13.2 & 37.0  \\
           &  (4.5)	 & (2.4) &	(4.0)	& (13.5)&	(24.7) & (10.9)	 \\
            \hline
        \multicolumn{8}{@{}c}{{Scenario 5}}\\
 & (.01;.10) & (.05;.35) & \underline{\textbf{(.15;.60)}} & \underline{(.20;.60)} & (.45;.60) & (.60;.60)\\
    ${\rm WE}$ &  0.1 & 6.2 & 60.3 & 28.9  & 3.6 &  0.8 & 0.1 & 7.9 &  29.8   \\
   &  (6.4) & (12.3) & (29.4)  & (10.5)  & (1.3)  & (0.1) \\
             ${\rm WE  (R)}$ &  0.1 & 7.3  &54.5 & 35.4 &  4.2 & 0.2 &0.3  &8.7 & 29.8  \\
         &    (6.1)  & (12.9) & (23.0) & (14.8)  & (3.0)  & (0.2) \\
   ${\rm MTA}$ &  0.0 &  8.5 & 55.3 & 25.3 &   9.7 & 1.2 &0.1 & 12.1 & 29.4  \\
   & (5.0) &  (8.7) &  (18.5) &  (15.6) &  (10.1) & (2.0) \\
           ${\rm WT}$ & 0.1 &	2.7 &	27.0	 &53.0 &	16.9 &	0.2 & 0.1 & 13.5 &32.3   \\
        &   (5.2) & 	(4.4)	 & (14.3) &	(22.3) &	(11.0)	 & (2.8)	\\
           \hline
                \end{tabular}}
\end{table}

\begin{table}
   \caption{Operating charsteristics of WE, WE(R), MTA and WT design in scenarios 6-10: recommendation proportions, mean number of patients assigned to a dose (in brackets), termination proportion (Term), mean number of toxicity (T) and efficacy (E) respones. The optimal dose is in \textbf{bold} and correct doses are \underline{underlined}. Results are based on $10^4$ replications. \label{tab:app2} }
   \fbox{%
\begin{tabular}{cccccccccc}
   & $d_1$ &   $d_2$ &  $d_3$ &   $d_4$ &   $d_5$ & $d_6$ & Term & T & E\\
              \hline
                \multicolumn{8}{@{}c}{{Scenario 6}}\\
  & (.01;.05) & (.05;.10) & (.10;.20) & (.20;.35) & \underline{\textbf{(.30;.55)}} & (.50;.55)\\
    ${\rm WE}$ & 0.4 &  0.8 & 3.7 & 18.9 & 53.5 & 18.9  & 4.7 & 11.0 & 19.4   \\
 &  (6.6) &  (7.4) & (10.0) & (15.7)  & (16.8)  & (2.8) \\
           ${\rm WE  (R)}$ & 0.4  & 0.8  & 4.8 & 25.4 & 56.7  & 7.1 &4.4 & 10.8 & 19.2  \\
          &  (6.3) &  (7.7)  &(11.1) & (16.3) & (15.2)  & (2.9) \\
   ${\rm MTA}$ &  0.1  & 0.7 & 4.5  & 17.0&  55.9 & 13.7 &8.3 & 13.2 &  21.4 \\
   & (4.5)  & (5.5) &  (7.9) &  (12.4) & (19.0) &   (7.8) \\
           ${\rm WT}$ &  0.2 &	0.9 &	3.8 &	21.4 &65.2	 & 5.8 & 2.7 & 14.7 & 24.4	   \\
          & (5.0) &	 (3.0)&	 	(5.3) &		(13.0) 	&	(25.1)& (7.9) \\
        \hline
     \multicolumn{8}{@{}c}{{Scenario 7}}\\
 & (.02;.30)& {(.07;.50)}& \underline{\textbf{(.13;.70)}}& \underline{(.17;.73)} & \underline{(.25;.76)} & \underline{(.30;.77)} \\
    ${\rm WE}$ &  0.5 &  21.4 &60.0 & 16.2 & 1.7  &0.0 & 0.0 & 5.9 & 34.6 \\
   &  (8.7)  &(20.6) & (25.3)  & (5.1)  & (0.4) & (0.0) \\
          ${\rm WE  (R)}$ & 0.8 & 10.9  & 48.5  & 31.9  & 7.2 &  0.6 & 0.0 & 6.9 &36.5 \\
        &   (8.6)  & (13.8) & (22.0) & (12.5)  &  (2.9)  & (0.3) \\
   ${\rm MTA}$ & 1.4 &8.7& 37.9& 24.5 &16.4 &11.1 &0.0 & 9.6 & 39.1 \\
  &  (6.2) &  (8.9) &  (14.6) &  (14.0) &  (11.3) &  (5.1) \\
           ${\rm WT}$ & 1.6 &	5.2	& 27.1&	29.8 &	24.7 &	11.7	  & 0.0 & 10.0 & 39.9 \\
        &   (6.7) &	(5.4)	 &(13.7)	&(14.8) &	(12.1)	 &(7.3)	 \\

   \hline
              \multicolumn{8}{@{}c}{{Scenario 8}}\\
 & (.03;.30) & \underline{\textbf{(.06;.50)}} & \underline{(.10;.52)} & \underline{(.20;.54)}& (.40;.55) & (.50;.55) \\
    ${\rm WE}$ &  3.2  & 37.8 &  34.7 & 19.1 &  4.2 & 1.0 &0.0 & 5.6 & 28.9  \\
   & (9.4) & (23.9) &  (17.8) &  (7.1)  & (1.5) & (0.1) \\
          ${\rm WE (R)}$ & 3.9 & 36.4 & 33.8 & 20.9 &  4.4  & 0.4 & 0.0 & 6.9 & 29.2  \\
          & (9.2) &  (16.9) & (18.7)  &(11.7)  & (3.2) &  (0.3) \\
   ${\rm MTA}$ &  12.8 &  43.0 &  21.7 & 12.7 & 8.2 & 1.7 & 0.1 & 9.3 &  29.3 \\
  &  (10.1) & (14.3)  & (12.8) &  (12.3) & (8.5) & (2.0) \\
           ${\rm WT}$ &  7.1 &	26.1 &	27.0 &	28.0	 &11.3 &	0.4 & 0.0 & 9.1 & 29.2  \\
            & (9.9)	&(13.0) &	(13.9)	 &(13.8) & 	(7.1) &	(2.2)	 \\
                   \hline
                        \multicolumn{8}{@{}c}{{Scenario 9}}\\
 & (.01;.30) & (.05;.50) & \underline{\textbf{(.10;.60)}} & (.15;.40) & (.20;.25) & (.25;.15)\\
    ${\rm WE}$ &   3.0 & 34.7  & 54.7 &  5.8  & 1.3   & 0.5 &0.0 &4.4 & 29.7 \\
    & (8.9)  & (23.1)  & (22.5)  & (3.9) &  (1.4) & (0.4) \\
           ${\rm WE  (R)}$ & 4.1 & 31.4  & 56.7 &  6.5  & 1.1  & 0.1 &0.0 & 5.5 & 28.4 \\
        &   (8.5)  &(15.6) & (22.0)  & (9.1)   & (3.6) & (1.2) \\
   ${\rm MTA}$ &  24.2 & 54.7 &  20.3 & 0.6 & 0.0& 0.0 &0.2 &5.0 &26.3  \\
  &  (13.2) &  (18.3) &  (14.1) &  (8.5) &  (4.7)  & (1.2) \\
  ${\rm WT}$ & 7.3 &	30.9 &	50.1 &	8.8 &	2.1 &	0.7	 & 0.2 & 5.5 & 27.9 \\
  &  (9.8)	& (15.4)	&  (22.1) &	(7.1) &	(3.3) &	(2.2)	 \\
                \hline
         \multicolumn{8}{@{}c}{{Scenario 10}}\\
 & (.02;.38) & \underline{\textbf{(.06;.50)}} & (.12;.40) & (.30;.30) & (.40;.25) & (.50;.20)\\
    ${\rm WE}$ &  18.4 & 55.9  &15.9 &  3.9  & 3.2   & 2.7  & 0.1 &5.7 & 25.6 \\
    & (14.7) &  (27.4)  & (11.0)  &  (4.3)  &  (2.1)  & (0.5) \\
           ${\rm WE  (R)}$ & 22.7 &  56.2 & 17.0  & 2.5  & 1.2  & 0.2 & 0.1 &6.7 & 24.6  \\
        &   (13.4) & (21.4) & (14.9) &  (6.6)  & (3.0) &  (0.6) \\
   ${\rm MTA}$ &  60.4&  35.3  & 2.8 & 0.4 & 0.1 & 0.1 &1.0 &6.3 & 23.9  \\
    & (20.2) & (17.6) &  (10.1) &  (8.0) &  (3.3) &  (0.5) \\
           ${\rm WT}$ &  29.0 &	49.3 &	15.8	 & 4.9 &	0.9	 & 0.1 &0.1 &5.9 & 24.8 \\
         &  (19.5) & 	(21.6)	 & (10.5) & 	(4.8)	& (2.3)	& (1.2)	 \\
      \hline      
      \end{tabular}}
\end{table}

\begin{table}
   \caption{Operating charsteristics of WE, WE(R), MTA and WT design in scenarios 11-14: recommendation proportions, mean number of patients assigned to a dose (in brackets), termination proportion (Term), mean number of toxicity (T) and efficacy (E) respones. The optimal dose is in \textbf{bold} and correct doses are \underline{underlined}. Results are based on $10^4$ replications. \label{tab:app3} }
  \fbox{%
\begin{tabular}{cccccccccc}
   & $d_1$ &   $d_2$ &  $d_3$ &   $d_4$ &   $d_5$ & $d_6$ & Term & T & E\\
          \hline
         \multicolumn{8}{@{}c}{{Scenario 11}}\\
         & (.03;.25) & (.09;.35) & (.16;.48) &  \underline{\textbf{(.28;.65)}} & (.42;.52) & (.56;.39) \\
         ${\rm WE}$ & 2.2 & 9.8  &30.4 & 46.5  & 8.6 & 2.8 & 0.2 &10.0 & 27.4 \\
        &  (9.3) & (13.5)  & (19.4) & (15.1) &  (2.4) & (0.3) \\
          ${\rm WE (R)}$ &  3.6 & 12.0 & 30.9  &47.9  & 5.3 &0.5 & 0.2  &10.0 & 27.1 \\
         & (10.2) & (13.9) & (18.0) & (14.5)  & (3.0)  & (0.2) \\
   ${\rm MTA}$ &  6.7 & 14.0 & 27.0 &  46.0 &  5.6 & 0.3 & 0.3 & 12.8 &  28.7 \\
  &  (8.1) &  (10.2) &  (14.1) &  (17.0) &  (8.7)  & (1.7) \\
              ${\rm WT}$ & 6.9	 & 9.4 & 	23.2 & 	56.9	 & 3.5 & 	0.0 & 0.1 &12.2 &  29.4 \\
    &        (9.9)	 & (7.9)	 &(13.8)	 &(22.2)	 &(4.6)	 &(1.6)	 \\
        \hline
              \multicolumn{8}{@{}c}{{Scenario 12}}\\
         & \underline{\textbf{(.02;.68)}}& (.05;.56) & (.07;.49) & (.09;.40) & (.11;.33) & (.13;.26) \\
   ${\rm WE}$ &   80.1 &  14.5 &  3.9 & 1.1  & 0.3  & 0.1 &0.0 & 1.8 & 35.4  \\
 &  (44.8) &  (10.0)  & (3.4) & (1.1) & (0.5) & (0.1) \\
      ${\rm WE (R)}$ & 70.9 & 18.7  & 7.1  & 2.4  & 0.9 & 0.2 &0.0 &3.3 &32.7   \\
      & (20.5)  & (15.1) &  (11.8)&   (7.0)  & (3.6)  & (1.9) \\
   ${\rm MTA}$ &  96.1  & 3.2 &  0.6  & 0.1 & 0.0 &  0.0 & 0.0 & 2.6 &35.5 \\
  &  (34.6) &  (10.1) &  (6.4) &  (4.8)  & (2.8) &  (1.4) \\
                  ${\rm WT}$ &  75.9	& 17.1 & 	5.1 & 	1.2 & 	0.6	 & 0.2 &0.0 &2.4 & 36.3 \\
             &     (37.0)	& (11.0)&	(5.4) &	(3.0) &	(2.1)	 & (1.4) \\
      \hline
             \multicolumn{8}{@{}c}{{Scenario 13}}\\
 & (.05;.01) & (.10;.02) & (.25;.05) & (.55;.35) & (.70;.55) & (.90;.70)\\
    ${\rm WE}$ &  0.0 & 0.0 & 0.2 & 2.9 & 0.6  &1.1& 95.2 & 11.7 & 6.0  \\
   &  (6.7)  & (7.6)  &(10.9)  &(12.2)  & (1.6)   & (0.1) \\
           ${\rm WE  (R)}$ & 0.1 & 0.0 & 0.2 & 4.4 & 1.1& 0.2 & 93.9 & 13.0 & 7.1   \\
         &  (6.9) &  (7.7) & (10.6)&  (12.1)&   (3.2)  & (0.4) \\
   ${\rm MTA}$ &  0.0 &  0.0 &  2.3  &  5.8&  0.0 & 0.0 & 91.9 & 11.0 &  6.2 \\
  &  (5.8) &  (5.9) & (7.7) &  (11.0) &  (2.7) & (0.3) \\
           ${\rm WT}$ & 	0.0 &	0.1	 & 5.5	& 3.1	& 0.0	& 0.0  &91.5 &11.2 & 5.4  \\
        &    (6.3)	 & (6.5)	 &(14.8)	 & (7.9)	& (1.5)&	(1.2)	\\
          \hline
        \multicolumn{8}{@{}c}{{Scenario 14}}\\
 & (.50;.40) & (.60;.55) & (.69;.65) & (.76;.65) & (.82;.65) & (.89;.65)\\
    ${\rm WE}$ &  2.2 & 0.2 & 0.1 & 0.3 & 0.2 & 0.0 & 96.9 & 11.0 & 9.1  \\
   &  (17.6)  & (2.8)  & (0.6)  & (0.1)  & (0.0) &  (0.0) \\
           ${\rm WE (R)}$ &  1.9 & 0.5  &0.3 & 0.2  &0.0  &0.0 & 97.4 &10.9 &8.9   \\
          & (17.6) &  (2.4)  & (0.8)  & (0.1)  &(0.0)  & (0.0) \\
   ${\rm MTA}$ & 8.8 & 0.2  & 0.0& 0.0  & 0.0 &  0.0 & 91.0 &11.5 & 9.6  \\
   &  (16.0) &  (4.3)  & (1.1) & (0.2) &  (0.0) &  (0.0) \\
           ${\rm WT}$ & 	9.5 &	0.0 &	0.0	 &0.0 &	0.0 & 	0.0	 & 90.4&12.1 & 9.7 \\
        &    (23.0) &	(0.4)	 &(0.1)	& (0.1) &	(0.1)	 & (0.1) \\
      \end{tabular}}
\end{table}

\end{document}